\documentclass[aip,amsmath,amssymb,reprint]{revtex4-1}

\usepackage{graphicx}% Include figure files
\usepackage{dcolumn}% Align table columns on decimal point
\usepackage{bm}% bold math
\usepackage[utf8]{inputenc}
\usepackage[T1]{fontenc}
\usepackage{mathptmx}

\usepackage{hyperref} %BS: This fixes some of the errors and fixes the links

\newcommand\mnras{{MNRAS}}     % Monthly Notices of the RAS 
\newcommand\aap{{A\&A}}        % Astronomy and Astrophysics
\newcommand\pasj{{PASJ}}       % Publications of the Astronomical Society of Japan
\newcommand\apjs{{ApJS}}       % Astrophysical Journal Supplement
\newcommand*\jgr{J. Geophys. Res}  %GIULIA, CHECK THIS!!!!!
\usepackage[dvipsnames]{xcolor}

\graphicspath{{./}{figures/}}

\begin{document}

\preprint{AIP/123-QED}

\title[Collisional ionisation and recombination effects on coalescence instability in chromospheric partially ionised plasmas]{Collisional ionisation and recombination effects on coalescence instability in chromospheric partially ionised plasmas}

\author{Giulia Murtas}
 \email{gm442@exeter.ac.uk}
 \affiliation{College of Engineering, Mathematics and Physical Sciences, Harrison Building, Streatham Campus, University of Exeter, North Park Road, Exeter, UK, EX4 4QF}
 
\author{Andrew Hillier}
\affiliation{College of Engineering, Mathematics and Physical Sciences, Harrison Building, Streatham Campus, University of Exeter, North Park Road, Exeter, UK, EX4 4QF}

\author{Ben Snow}
\affiliation{College of Engineering, Mathematics and Physical Sciences, Harrison Building, Streatham Campus, University of Exeter, North Park Road, Exeter, UK, EX4 4QF}

\date{\today}

\begin{abstract}

Plasmoid-mediated fast magnetic reconnection plays a fundamental role in driving explosive dynamics and heating, but relatively little is known about how it develops in partially ionised plasmas (PIP) of the solar chromosphere. Partial ionisation might largely alter the dynamics of the coalescence instability, which promotes fast reconnection and forms a turbulent reconnecting current sheet through plasmoid interaction, but it is still unclear to what extent PIP effects influence this process. We investigate the role of collisional ionisation and recombination in the development of plasmoid coalescence in PIP through 2.5D simulations of a two-fluid model. The aim is to understand whether these two-fluid coupling processes play a role in accelerating reconnection. We find that in general ionisation-recombination process slow down the coalescence. Unlike the previous models in G. Murtas, A. Hillier \& B. Snow, Physics of Plasmas 28, 032901 (2021) that included thermal collisions only, ionisation and recombination stabilise current sheets and suppress non-linear dynamics, with turbulent reconnection occurring in limited cases: bursts of ionisation lead to the formation of thicker current sheets, even when radiative losses are included to cool the system. Therefore, the coalescence time scale is very sensitive to ionisation-recombination processes. However, reconnection in PIP is still faster than in a fully ionised plasma environment having the same bulk density: the PIP reconnection rate ($M_{_{\operatorname{IRIP}}} = 0.057$) increases by a factor of $\sim 1.2$ with respect to the MHD reconnection rate ($M_{_{\operatorname{MHD}}} = 0.047$).

\end{abstract}

\maketitle

\section{Introduction}
\label{sec:intro}

Magnetic reconnection takes place in a wide range of astrophysical settings \citep{doi:10.1146/annurev-astro-082708-101726}: it occurs in presence of a non negligible resistivity, when magnetic field lines change their connectivity altering the topology of the magnetic field. During reconnection, the frozen-in constraint imposed by ideal magnetohydrodynamics (MHD) no longer applies and the energy released by reconnection energises particles to high energies and heats the plasma \citep{2000mrp..book.....B,2000mrmt.conf.....P}. Explosive events in the solar chromosphere, such as chromospheric jets \citep{2007Sci...318.1591S,2011ApJ...731...43N} and Ellerman bombs \citep{1917ApJ....46..298E,2018PhPl...25d2903N}, are believed to be driven by fast magnetic reconnection, responsible of efficiently releasing the stored magnetic energy into thermal and kinetic energy\citep{2000mrp..book.....B,2000mrmt.conf.....P} at shorter time scales than the classical reconnection models.

The hypothesis of fast magnetic reconnection as a driver of explosive phenomena in chromospheric plasmas has been reinforced by the identification of plasma blobs in the outflow of chromospheric jets and UV bursts\citep{2011PhPl...18k1210S,2012ApJ...759...33S,Guo_2020}, colliding and merging with each other before being ejected from the current sheet. These structures are generally interpreted to be plasmoids, concentrations of current density trapped in closed loops of magnetic field lines that are commonly present in reconnecting systems \citep{1963PhFl....6..459F,2001ApJ...551..312T,2001ApJ...551..312T,2001EP&S...53..473S,2009PhRvL.103j5004S,2007PhPl...14j0703L,2012PhPl...19d2303L,2016PPCF...58a4021L}. Plasmoids are believed to play a major role in speeding up reconnection to the time scales found in observations, as their formation directly affects the current sheet size \citep{1989ApJ...340..550Z}. Plasmoids can be the result of the tearing instability, which breaks thin current sheets (where the current sheet aspect ratio $\delta / L \ll 1$) into fragments\citep{1963PhFl....6..459F,2007PhPl...14j0703L,2015ApJ...807..159T}. The resulting high current densities in each of these fragments facilitate a high reconnection rate \citep{2015ApJ...799...79N}.

Plasmoid formation due to the instability of current sheets has been extensively examined through numerical studies \citep{1984PhFl...27..137P,1984PhFl...27.1207S,1986PhFl...29.1520B,1986JGR....91.6807L,1991PhFlB...3.1927J,1995PhPl....2..388U,2005PhRvL..95w5003L,2016PPCF...58a4021L}. This mechanism is dependent on the value of the Lundquist number\citep{1987ApJ...321.1031T, 2001EP&S...53..473S,2011ApJ...730...47B}, $S = \frac{L v_A}{\eta}$, where $L$ is a characteristic length of the system, $v_A$ is the Alfv\'en speed and $\eta$ is the diffusivity. Several works proved that it is possible to speed up reconnection in thin current sheets for a critical value of the Lundquist number typically $\sim 10^3 - 10^4$ for fully ionised plasmas \citep{2005PhRvL..95w5003L,2009PhRvL.103j5004S,2009PhPl...16k2102B,2009PhPl...16l0702C,2010PhPl...17f2104H,2012PhPl...19d2303L,2010PhPl...17e2109N,2012PhPl...19g2902N,2013PhPl...20f1206N,2015ApJ...799...79N,2016PPCF...58a4021L}. Above this limit, current sheets become unstable and plasmoids formation occurs.

Once plasmoids are generated, they are pulled against each other and merge, in a mechanism that further increase the reconnection rate: this process is called coalescence instability \citep{1986ITPS...14..929T,2000mrp..book.....B}. Plasmoid coalescence occurs in the nonlinear tearing mode phase between plasmoids sharing an $X$-point and evolves through two separate phases. The first is an ideal MHD phase with a growth rate that is almost independent of $\eta$ \citep{1986ITPS...14..929T}. In this phase, the two plasmoids move close and a current sheet forms between them. The second part of coalescence evolves in a resistive phase, where the current sheet begins to reconnect \citep{2000mrp..book.....B}. As further plasmoids are formed in between the coalescing plasmoids, this instability can become a fractal process operating at multi-spatial scale, depending on the Lundquist number \citep{1987ApJ...321.1031T,2001EP&S...53..473S,2011ApJ...730...47B}. 

A large portion of plasmas across the universe are only partially ionised, for example the chromospheric plasma, whose ionisation degree falls in the range $10^{-4} - 10^{-1}$, as reported by many studies\citep{1981ApJS...45..635V,1986A&A...154..231P,2008ApJS..175..229A,Khomenko2008,2015ApJ...799...79N}. The presence of neutral species might alter the plasmoid dynamics, as further physical processes develop from the coupling with charged particles. The role of partial ionisation on the onset of magnetic reconnection and resistive tearing instability was investigated in several studies \citep{1989ApJ...340..550Z,2011PhPl...18k1210S,2011PhPl...18k1211Z,2012ApJ...760..109L,2013PhPl...20f1202L,2015ApJ...799...79N,2015PASJ...67...96S}. It has been observed that partial ionisation largely modifies the reconnection rate by changing the Alfv\'en speed, which in turn affects the Lundquist number of the system\citep{2012ApJ...760..109L} and the conditions for plasmoids formation.

The physics of a reconnecting current sheet is also affected by processes of collisional ionisation\citep{1997ADNDT..65....1V} and recombination, which actively modify the relative abundance of the plasma species. As pointed out from many studies of magnetic reconnection in a multi-fluid partially ionised plasma at low $\beta$ \citep{2018ApJ...868..144N,2012ApJ...760..109L,2013PhPl...20f1202L,2015ApJ...805..134M,2018ApJ...852...95N,2018PhPl...25d2903N} the non-equilibrium ionisation-recombination leads to a strong ionisation of the material in the reconnection region and a faster reconnection rate developing before the onset of plasmoid instabilities\citep{2018ApJ...852...95N}. Past studies \citep{2013PhPl...20f1202L} reported an increase of the ionisation degree by an order of magnitude within the current sheet during reconnection. The strong ionisation is responsible for a larger interaction between the neutral fluid and the plasma, with a stronger coupling occurring both in the inflow and outflow region. In case of a plasma $\beta$ smaller than 1, however, plasmoid instability remains the main process promoting fast magnetic reconnection\citep{2018ApJ...852...95N}.

The ionisation-recombination effects are further enhanced by the action of the ionisation potential. When collisional ionisation takes place,
the energy expended by a free electron to release a bound electron results in a net loss of energy from the plasma\citep{Snow2021}. As the recombination process is associated with changes in energy levels and photons being released, this overall effect can be modeled as a radiative loss. Studies investigating the role of radiative cooling in magnetic reconnection \citep{1995ApJ...449..777D,2011PhPl...18d2105U} proved that the inclusion of the ionisation potential thins the reconnection layer by decreasing the plasma pressure and density inside the current sheet. Therefore, adding radiative losses speeds up reconnection to higher rates than the ones of models without radiation and might lead to time scales and outflows that are consistent with those found in spicules and chromospheric jets \citep{2017ApJ...842..117A}.

Relatively little is known about how the coalescence instability is altered by the multi-fluid physics of a partially ionised plasma. In our preliminary study \citep{doi:10.1063/5.0032236} we found that, in an ion-neutral plasma with fluids coupled through elastic collisions, partial ionisation speeds up both phases of coalescence and promotes non-linear effects during reconnection. In this paper we examine in detail the effects of collisional ionisation, recombination and optically thin radiative losses on the coalescence instability developing in partially ionised plasmas. This model improves upon our previous work \citep{doi:10.1063/5.0032236} by including the contribution of these processes. We aim to understand how the new type of coupling between charges and neutral species influence the reconnection rate, compared to our previous research. In Section \ref{sec:methods} we discuss the two-fluid model employed for our simulations. In Section \ref{sec:2.5D_cases} we report the results of our 2.5D simulations. The results are discussed in Section \ref{sec:discussion}.

\section{Methods}
\label{sec:methods}

Numerical simulations are performed of the coalescence instability in 2.5D using the (P\underline{I}P) code \citep{2016A&A...591A.112H}. The process is studied in single-fluid fully ionised cases (MHD) and two-fluid partially ionised cases (PIP). MHD cases are modelled by a charge-neutral ion-electron hydrogen plasma, while PIP cases are characterised by a neutral fluid and a hydrogen plasma being collisionally coupled. The fluids are described by two separate sets of non-dimensional equations \citep{2019PhPl...26h2902H}, which have been derived from previous models \cite{1965RvPP....1..205B,2012ApJ...760..109L,2012PhPl...19g2508M}. The equations are solved throughout the domain through a fourth-order central difference scheme, and the physical variable updates are computed by a four-step Runge-Kutta\cite{2005A&A...429..335V} scheme for time integration. The neutral fluid is governed by compressible inviscid hydrodynamics equations:
\begin{equation}
    \frac{\partial \rho_n}{\partial t} + \nabla \cdot (\rho_n \textbf{v}_n) = D,
\end{equation}
\begin{equation}
    \frac{\partial}{\partial t}(\rho_n \textbf{v}_n) + \nabla \cdot (\rho_n \textbf{v}_n \textbf{v}_n + p_n \textbf{I}) = \mathbf{R},
    \label{eq:force_neutral}
\end{equation}
\begin{equation}
    \frac{\partial e_n}{\partial t}  + \nabla \cdot [\textbf{v}_n (e_n + p_n)] = E,
    \label{eq:neutral_energy_2}
\end{equation}
\begin{equation}
    e_n = \frac{p_n}{\gamma -1} + \frac{1}{2} \rho_n v_{n}^{2},
    \label{neutral_energy}
\end{equation}
\begin{equation}
    T_n = \gamma \frac{p_n}{\rho_n},
    \label{eq:neutral_temperature}
\end{equation}
while compressible inviscid resistive MHD equations model the plasma:
\begin{equation}
    \frac{\partial \rho_p}{\partial t} + \nabla \cdot (\rho_p \textbf{v}_p) = - D,
\end{equation}
\begin{equation}
    \frac{\partial}{\partial t}(\rho_p \textbf{v}_p) + \nabla \cdot \Bigg(\rho_p \textbf{v}_p \textbf{v}_p + p_p \textbf{I} - \textbf{B} \textbf{B} +  \frac{\textbf{B}^2}{2} \textbf{I} \Bigg) = - \mathbf{R},
    \label{eq:force_plasma}
\end{equation}
\begin{multline}
    \frac{\partial}{\partial t} \Bigg( e_p + \frac{B^2}{2}\Bigg) + \nabla \cdot [ \textbf{v}_p (e_p + p_p) \\
    -(\textbf{v}_p \times \textbf{B}) \times \textbf{B} + \eta (\nabla \times \textbf{B}) \times \textbf{B}] = -E - \Phi_I + A_{heat},
    \label{eq:plasma_energy_equation}
\end{multline}
\begin{equation}
    \frac{\partial \textbf{B}}{\partial t} - \nabla \times (\textbf{v}_p \times \textbf{B} - \eta \nabla \times \textbf{B}) = 0,
\end{equation}
\begin{equation}
    e_p = \frac{p_p}{\gamma -1} + \frac{1}{2} \rho_p v_{p}^{2},
    \label{plasma_energy}
\end{equation}
\begin{equation}
    \nabla \cdot \textbf{B} = 0,
\end{equation}
\begin{equation}
    T_p = \gamma \frac{p_p}{2\rho_p}.
    \label{eq:plasma_temperature}
\end{equation}
The subscripts $p$ and $n$ in both plasma and neutral equations identify physical quantities of the ion-electron plasma and the neutral fluid respectively. The variables $\mathbf{v}$, $p$, $\rho$, $T$ and $e$ are respectively the fluids velocity, gas pressure, density, temperature and internal energy, $\gamma= 5/3$ is the adiabatic index and \textbf{B} is the magnetic field. The terms $D$, $\mathbf{R}$ and $E$ are respectively the source terms for mass, momentum and energy transfer between the two species, and are defined as follows:
\begin{equation}
    D = \Gamma_{rec} \rho_p - \Gamma_{ion} \rho_n,
    \label{eq:mass_source_term}
\end{equation}
\begin{equation}
    \mathbf{R} = - \alpha_c \rho_n \rho_p (\textbf{v}_n - \textbf{v}_p ) + \Gamma_{rec} \rho_p \mathbf{v}_p - \Gamma_{ion} \rho_n \mathbf{v}_n,
\end{equation}
\begin{multline}
    E = - \alpha_c \rho_n \rho_p \Bigg[ \frac{1}{2} (\textbf{v}_n ^2 -\textbf{v}_p ^2) + \frac{T_n - T_p}{\gamma (\gamma -1)} \Bigg] \\ +\frac{1}{2} (\Gamma_{rec} \rho_p \mathbf{v}_p ^2 - \Gamma_{ion} \rho_n \mathbf{v}_n ^2) + \frac{\Gamma_{rec} \rho_p T_p - \Gamma_{ion} \rho_n T_n}{\gamma (\gamma -1)}.
\end{multline}
Both fluids are subject to the ideal gas law. The factor of 2 in Equation (\ref{eq:plasma_temperature}) is included to account for the electron pressure in the plasma contribution.

The two-fluid collisional coupling is determined by the parameter $\alpha_c (T_n , T_p ,v_D)$, whose non-dimensional expression\cite{1986MNRAS.220..133D}, that includes charge exchange\cite{2018ApJ...869...23Z}, is found in Equation (\ref{alpha_c}):
\begin{equation}
\alpha_c = \alpha_c (0) \sqrt{\frac{T_n +T_p}{2}} \sqrt{1 + \frac{9\pi}{64} \frac{\gamma}{2(T_n + T_p)} v_D^2},
\label{alpha_c}
\end{equation}
where $\alpha_c (0)$ is the initial coupling and $v_D$ = $\mid$ \textbf{v$_n$} - \textbf{v$_p$} $\mid$ is the magnitude of the drift velocity between the neutral components and the hydrogen plasma. When the magnitude of the drift velocity becomes bigger than the thermal velocity, the particles are subject to a higher number of collisions as they are drifting past each other: the expression for $\alpha_c$ in Equation (\ref{alpha_c}) takes into account the higher number of collisions occurring at supersonic drift velocities. The collisional coupling between ions and electrons is modeled by imposing a spatially uniform, constant diffusivity ($\eta$) in the system. The two fluids are in an initial thermal equilibrium, but there is not an imposed initial ionisation equilibrium. The Hall effect is not included in this study.

The terms $\Gamma_{rec}$ and $\Gamma_{ion}$ are the recombination and collisional ionisation rates for a hydrogen atom. The normalised empirical forms of the rates \cite{1997ADNDT..65....1V,2003poai.book.....S} are:
\begin{equation}
    \Gamma_{rec} = \frac{\rho_p}{\sqrt{T_p}} \frac{\sqrt{T_f}}{\xi_{p}(0)} \tau_{_{\operatorname{IR}}},
    \label{eq:recombination_rate}
\end{equation}
\begin{equation}
    \Gamma_{ion} = \rho_p \Bigg( \frac{e^{- \chi} \chi^{0.39}}{0.232 + \chi} \Bigg) \frac{\hat{R}}{\xi_{p}(0)} \tau_{_{\operatorname{IR}}},
    \label{eq:ionisation_rate}
\end{equation}
where:
\begin{equation}
    \chi = 13.6 \frac{T_f}{T_{e0} T_p},
\end{equation}
\begin{equation}
    \hat{R} = \frac{2.91 \cdot 10^{-14}}{2.6 \cdot 10^{-19}} \sqrt{T_{e0}}.
    \label{eq:R_hat}
\end{equation}
The rates of collisional ionisation used in this work are based on a semi-empirical model for the ionisation of hydrogen by electron impact that assumes ionisation from the ground state\citep{1983JPCRD..12..891B,1988JPCRD..17.1285L,1997ADNDT..65....1V}. These rates do not include photo-ionisation or ionisation from excited states, which are known to be important for the chromosphere\citep{1985irss.rept..213G,1989A&A...225..222V,1998A&A...333.1069P}.

Two characteristic temperatures appear in Equations (\ref{eq:recombination_rate})-(\ref{eq:R_hat}), and are based on a physical reference electron temperature $T_0$ in K. $T_{e0}$ is the value of $T_0$ converted in electron volts. The initial normalisation of the system gives a bulk sound speed of unity. However, in the two-fluid model, this does not necessarily equate to a plasma temperature of unity. The ionisation and recombination rates depend on the electron temperature (which is assumed to be equal to the plasma temperature in our model), therefore a correction factor $T_f$ is needed to ensure that the desired dimensional electron temperature is being used to calculate the rates. For our normalisation, the plasma temperature is defined as:
\begin{multline}
T_p= \frac{\gamma p_p}{2 \rho_p} = \frac{\gamma}{2} \frac{2 \xi_p}{(\xi_n + 2 \xi_p)} \frac{p_p + p_n}{\xi_p (\rho_p+\rho_n)} \\
= \frac{\gamma (p_p + p_n)}{\rho_p+\rho_n} \frac{1}{\xi_n + 2 \xi_p} = \frac{ \gamma (p_p + p_n)}{\rho_p+\rho_n} \frac{1}{T_f}.
\end{multline}
where $\gamma(p_p + p_n)/(\rho_p+\rho_n)$ is the bulk sound speed squared and is initially equal to 1. Therefore, the factor $T_f$ is applied to ensure that the electron (plasma) temperature used in the ionisation-recombination rates corresponds initially to the reference dimensional temperature $T_0$.

The initial ion fraction $\xi_p$ is determined from the initial choice of the temperature $T_0$ through its relation with ionisation and recombination rates. In a steady state, the source term for mass $D$ in Equation (\ref{eq:mass_source_term}) is zero, leading to the relation:
\begin{equation}
    \Gamma_{rec} \rho_p = \Gamma_{ion} \rho_n.
    \label{eq:ion_fractio_start}
\end{equation}
Rearranging the terms in Equation (\ref{eq:ion_fractio_start}) we obtain:
\begin{equation}
    \xi_p = \frac{1}{1 + \frac{\Gamma_{rec}}{\Gamma_{ion}}}.
\end{equation}
Ionisation and recombination rates can be expressed as a function of the temperature $T$ and the plasma density $\rho_p$:
\begin{equation}
    \Gamma_{ion} = G(T) \rho_p,
\end{equation}
\begin{equation}
    \Gamma_{rec} = F(T) \rho_p,
\end{equation}
therefore the ion fraction can be also expressed as a function of the temperature:
\begin{equation}
    \xi_p = \frac{1}{1 + \frac{F(T)}{G(T)}}.
\end{equation}
In case of simulations not including ionisation-recombination processes, $\xi_p$ is imposed as an initial condition.

The free parameter $\tau_{_{\operatorname{IR}}}$ determines the relation of the recombination time scale with the dynamic time scale of the simulation, calculated from a characteristic length and a characteristic speed of the system (see later in the section for more details). In the (P\underline{I}P) code $\Gamma_{ion}$ and $\Gamma_{rec}$ are calculated in dimensional form from Equations (\ref{eq:recombination_rate}) and (\ref{eq:ionisation_rate}), then normalised by the recombination rate in order to obtain an initial $\Gamma_{rec} = 1$. The parameter $\tau_{_{\operatorname{IR}}}$ is imposed to re-scale $\Gamma_{rec}$, and fix an initial recombination time scale. For example, if $\tau_{_{\operatorname{IR}}}$ is set equal to $10^{-3}$ then the initial $\Gamma_{rec} = 10^{-3}$, and recombination would occur over a time scale of $10^3$. The ionisation rate $\Gamma_{ion}$ is also normalised by the same $\tau_{_{\operatorname{IR}}}$.

The terms $\Phi_I$ and $A_{heat}$ in Equation (\ref{eq:plasma_energy_equation}) are associated to the ionisation potential and account for radiative losses. $\Phi_I$ approximates the energy removed by the system through ionisation, while $A_{heat}$ is an arbitrary heating term included to obtain an initial equilibrium. Their non-dimensional forms are:
\begin{equation}
    \Phi_I = \Gamma_{ion} \rho_n \hat{\Phi},
    \label{eq:Phi_I}
\end{equation}
\begin{equation}
    A_{heat} = \Gamma_{ion} (t = 0) \rho_n (t = 0) \hat{\Phi},
    \label{eq:A_heat}
\end{equation}
where $\hat{\Phi} = 13.6 / (K_B \gamma T_0)$ ensures consistency between the normalisation of ionisation potential and the equations modelling the system. Here the Boltzmann constant is $K_B = 8.617 \cdot 10^{-5}$ eV K$^{-1}$. More details on the atomic internal structure used to estimate the ionisation potential can be found in a recent paper\cite{Snow2021}.

Both sets of equations are non-dimensionalised \citep{2019PhPl...26h2902H} by a reference density $\rho_0$ and the total sound speed $c_s = \sqrt{\gamma (p_n + p_p )/(\rho_n + \rho_p)}$, initially set equal to 1. For fully ionised plasmas (MHD cases), the initial density and pressure are set constant, uniform across the domain and equal to:
\begin{equation}
    \rho_{_{_{\operatorname{MHD}}}} = \xi_p \rho_0 = 1,
\end{equation}
\begin{equation}
    p_{_{\operatorname{MHD}}} = p_0 = \gamma^{-1},
\end{equation}
where $\xi_p$ is the ion fraction, equal to 1 for fully ionised plasmas. The cases that are run in a partially ionised plasma (PIP cases) present a similar normalisation as the one for a fully ionised plasma, where the bulk physical variables are set equal to the MHD values and uniform across the domain:
\begin{equation}
    \rho_n + \rho_p = \xi_n \rho_0 + \xi_p \rho_0 = 1,
\end{equation}
\begin{equation}
    p_n + p_p =  \frac{\xi_n}{(\xi_n + 2 \xi_p)} p_0 + \frac{2 \xi_p}{(\xi_n + 2 \xi_p)} p_0 = \gamma^{-1},
\end{equation}
where $\xi_n$ is the neutral fraction.

This normalisation is chosen to have the physical properties dependent on characteristic length scales, which are directly comparable to the size of the plasmoids involved in the merging. The non-dimensional collision frequency can be compared to the chromospheric dimensional values by dividing it by a characteristic dimensional time scale $\tau_{col}$. The parameter $\tau_{col}$ is found from the ratio of the physical plasmoid size (see Section \ref{subsec:initial_conditions} for more details) and a characteristic speed in the solar chromosphere, which in this study is chosen to be the sound speed ($\sim 10$ km s$^{-1}$). Therefore, the model can be easily re-scaled to examine the properties of magnetic structures of various sizes in the solar chromosphere, from a few meters to a few hundred kilometers.

\subsection{Initial conditions}
\label{subsec:initial_conditions}

\begin{figure}[htb]
    \centering
    \includegraphics[width=\columnwidth,clip=true,trim=0.5cm 2cm 1cm 3.6cm]{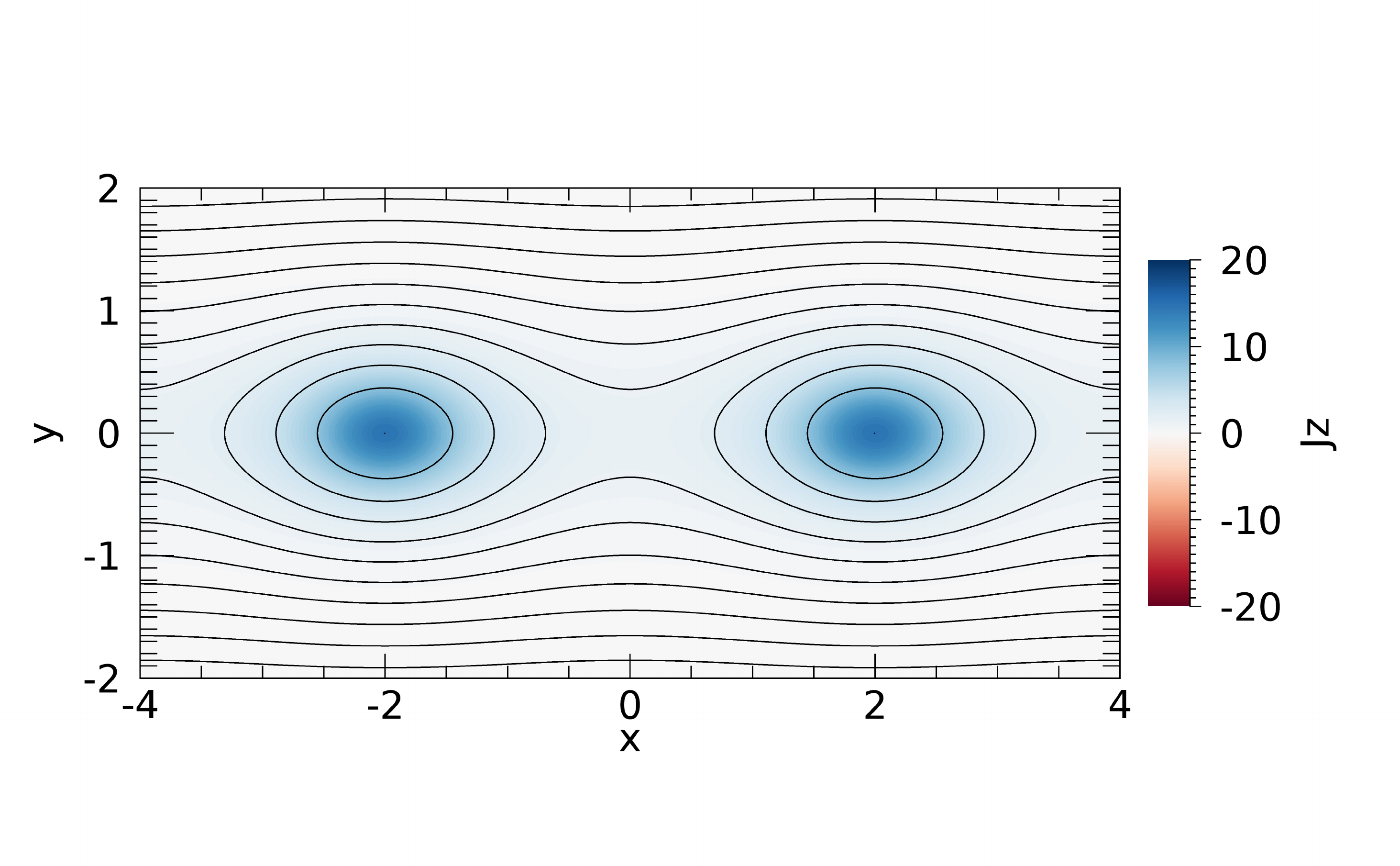}  
    \caption{Initial distribution for the current density $J_z$ ($t = 0$). The centre of the two initial plasmoids (blue spots) is located on the $x$-axis. Magnetic field lines are displayed in black.}
    \label{fig:initial_conditions}
\end{figure}

The initial setup of plasmoid coalescence is provided by a modified force-free Fadeev equilibrium \citep{Fadeev_1965,2000mrp..book.....B}, with the magnetic field components given by Equations (\ref{fadeev_bx_1})-(\ref{fadeev_bz_1}). Despite the photospheric magnetic field not being force-free at the boundary with the convective zone, its structure is rearranged before reaching the corona as the non force-free components decay due to the action of chromospheric neutrals \citep{2009ApJ...705.1183A}. For this reason, it is useful to impose an initial force-free field condition to the system. The classic Fadeev equilibrium magnetic field does not satisfy the condition $\textbf{J} \times \textbf{B} = 0$ for a force-free field. Therefore the traditional Fadeev equilibrium is modified by including a component $B_z$:
\begin{equation}
    B_{x} = - \frac{\sqrt{2 \gamma^{-1} \beta^{-1}} \epsilon \sin(kx)}{\cosh(ky) + \epsilon \cos(kx)},
    \label{fadeev_bx_1}
\end{equation}
\begin{equation}
    B_{y} = - \frac{\sqrt{2 \gamma^{-1} \beta^{-1}} \sinh(ky)}{\cosh(ky) + \epsilon \cos(kx)},
    \label{fadeev_by_1}
\end{equation}
\begin{equation}
    B_{z} = \frac{\sqrt{2 \gamma^{-1} \beta^{-1}} \sqrt{1- \epsilon^2}}{\cosh(ky) + \epsilon \cos(kx)}.
    \label{fadeev_bz_1}
\end{equation}
In the equations above, $k = \frac{\pi}{2}$ and $\epsilon = 0.5$, which lead to a moderately peaked current localisation at each plasmoid centre. In the limit for $\epsilon \rightarrow 0$ there is a less peaked current density and a weaker attraction between the plasmoids. When $\epsilon = 0$, $B_y$ reduces to a current sheet characterised by the tanh profile of the well known Harris sheet \citep{1962NCim...23..115H}. The limit $\epsilon \rightarrow 1$ corresponds to a peaked localization and stronger attraction forces. At the upper limit ($\epsilon = 1$), the current distribution becomes the delta function. The bulk (plasma + neutrals) plasma $\beta$, defined as $2(p_p + p_n)/B^2$, is set equal to 0.1 for all simulations. In the two-fluid cases it is possible to define a second plasma $\beta$ associated to the isolated plasma $2 p_p/B^2$, which from the initial condition is equal to 0.002. The initial current density distribution and magnetic field lines are shown in Figure \ref{fig:initial_conditions}. Note that these initial conditions mimic our previous paper\cite{doi:10.1063/5.0032236}. 

The Fadeev equilibrium is unstable to the coalescence instability \citep{2000mrp..book.....B}. An initial velocity perturbation in both plasma and neutral species is imposed to break the initial equilibrium, given by:
\begin{equation}
    v_{x,p} = v_{x,n} = -0.05 \sin \Bigg(\frac{kx}{2} \Bigg) e^{-y^2} + v_{\operatorname{noise}},
    \label{velocity_perturbation}
\end{equation}
where $v_{\operatorname{noise}}$ is a white noise component two orders of magnitude smaller than the main perturbation, included to simulate small environmental perturbations. The smaller magnitude of the white noise prevents it from dominating the motion of the two plasmoids during coalescence, while still promoting smaller scale dynamics by breaking the symmetry of the system. All simulations were performed with the same random noise seed. The sin$(x)$ term promotes attraction of the plasmoids, while the term dependent on $y$ localises the perturbation to a small region around the $x$-axis. Further details on the chosen velocity perturbation can be found in our recent paper \cite{doi:10.1063/5.0032236}.

All simulations are resolved by $4100 \times 3074$ grid cells, corresponding to a cell size of $\Delta x = 1.95 \cdot 10^{-3}$ and $\Delta y = 2.6 \cdot 10^{-3}$. The resolution has been tested in order to ensure that the current sheets are resolved with our grid. The initial separation between the plasmoids, calculated from $O$-point to $O$-point, (the centre of the blue spots in Figure \ref{fig:initial_conditions}) is equal to $4 L$, where $L$ is resolved by 513 grid points. The initial plasmoids width, calculated as the distance between top and bottom edges of the separatrix, is $1.66 L$ and determined by the initial magnetic field conditions.

The dynamics of the plasmoids merging is evaluated in a squared computational domain of $x = [-4, 4]$ and $ y = [-4, 4]$. We use symmetric boundaries at $y = -4$ and $y = 4$: in this configuration $v_x$ and $B_y$ change sign across the boundary, while $v_y$ and $B_x$ keep the same magnitude and sign across the boundary. The boundaries at $x = -4$ and $x = 4$ are chosen to be periodic.

\begin{figure*}[htb]
    \centering
    \includegraphics[width=\textwidth,clip=true,trim=0cm 1.5cm 0cm 2cm]{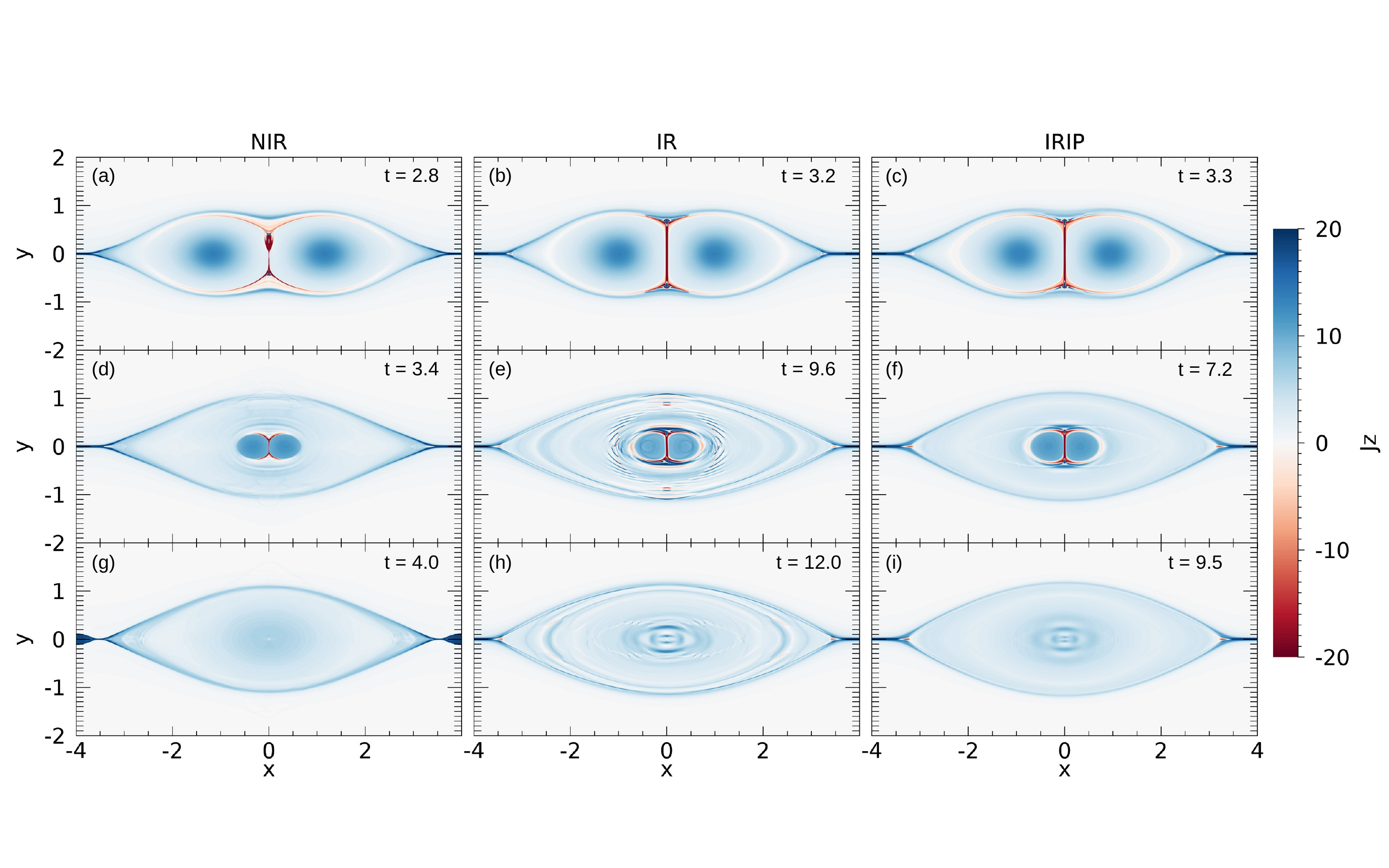}  
        \caption{Comparison of $J_z$ between three PIP cases with different two-fluid coupling processes: NIR (A1, left column), IR (A2, central column) and IRIP (A3, right column). The frames identify different steps of the coalescence instability. Panels (a), (b) and (c) show the initiation of the reconnection process. In panels (d), (e) and (f) the evolution of coalescence is displayed at later stages. The final stage of coalescence is shown in panels (g), (h), and (i) with the formation of the resulting plasmoid. Times are given in the same non-dimensional unit.}
    \label{fig:contour_Jz}
\end{figure*}

\section{Results}
\label{sec:2.5D_cases}

The coalescence instability has been well studied in MHD\citep{Finn1977,1986ITPS...14..929T,2000mrp..book.....B,Dorelli2001,PhysRevLett.96.135001,Makwana2018}. The general behaviour is that neighbouring plasmoids inside a current sheet attract each other, a second current sheet forms in between them and reconnection occurs in it, leading to the plasmoids complete merging. Extending the model to PIP (using only thermal collisions) results in a faster coalescence and the promotion of small-scale dynamics such as sub-critical secondary plasmoids formation and fractal coalescence (see our recent paper \citep{doi:10.1063/5.0032236}). Here the model is further extended to account for ionisation/recombination and ionisation potential effects, which provide a more realistic description of the physical processes occurring in a partially ionised plasma.

Initially, results are presented for weakly-coupled PIP cases with elastic collisions only (NIR), ionisation and recombination (IR), and ionisation, recombination and ionisation potential (IRIP), which are compared to a fiducial MHD simulation. Figure \ref{fig:contour_Jz} shows the time evolution of the coalescence through the current density for reference cases A1 (NIR), A2 (IR) and A3 (IRIP), corresponding to the three models for PIP. From this figure, we see major differences between NIR, IR and IRIP cases in time scale, reconnection rate and dynamics of each phase of coalescence. These results are discussed in detail in Section \ref{sec:reference_cases}. A parameter study is then performed in Section \ref{sec:tau_IR} to investigate the effect of varying $\tau_{_{\operatorname{IR}}}$, which determines the relative importance of ionisation and recombination rates to the collision rates, and on the initial ion fraction in Section \ref{sec:ion_fraction}. The full array of simulations is shown in Table \ref{tab:parameters_2D}. Simulations listed with letter M are run in fully ionised plasmas (MHD). Cases listed with letter A are the reference PIP simulations for the comparison of PIP coupling models. Cases listed with letter B compose the survey on $\tau_{_{\operatorname{IR}}}$. Cases listed with letter C compose the survey on the initial ion fraction $\xi_p (0)$.

\begin{table}
\caption{List of simulations parameters. \label{tab:parameters_2D}}
\begin{ruledtabular}
\begin{tabular}{ccccccc}
 ID & Model & $\eta$ & $\alpha_c$ (0) & $\xi_p$\footnotemark[2] (0) & $T_0$ (K)\footnotemark[2] & $\tau_{_{\operatorname{IR}}}$ \\
\hline
 M1 & MHD & $1.5 \cdot 10^{-3}$ & $\infty$\footnotemark[1] & 1\footnotemark[1] & - & -  \\ 
 M2 & MHD & $3 \cdot 10^{-3}$ & $\infty$\footnotemark[1] & 1\footnotemark[1] & - & -  \\
 \hline
 A1 & NIR & $1.5 \cdot 10^{-3}$ & $10^{2}$ & $10^{-2}$ & 10855 & - \\
 A2 & IR & $1.5 \cdot 10^{-3}$ & $10^{2}$ & $10^{-2}$ & 10855 & $10^{-3}$ \\
 A3 & IRIP & $1.5 \cdot 10^{-3}$ & $10^{2}$ & $10^{-2}$ & 10855 & $10^{-3}$ \\
 \hline
 B1 & IRIP & $1.5 \cdot 10^{-3}$ & 5 & $10^{-2}$ & 10855 & $5 \cdot 10^{-6}$ \\
 B2 & IRIP & $1.5 \cdot 10^{-3}$ & 5 & $10^{-2}$ & 10855 & $5 \cdot 10^{-5}$ \\
 B3 & IRIP & $1.5 \cdot 10^{-3}$ & 5 & $10^{-2}$ & 10855 & $5 \cdot 10^{-4}$ \\
 B4 & IRIP & $1.5 \cdot 10^{-3}$ & 5 & $10^{-2}$ & 10855 & $5 \cdot 10^{-3}$ \\
 B5 & IRIP & $1.5 \cdot 10^{-3}$ & 5 & $10^{-2}$ & 10855 & $5 \cdot 10^{-2}$ \\
 B6 & IRIP & $1.5 \cdot 10^{-3}$ & 5 & $10^{-2}$ & 10855 & $5 \cdot 10^{-1}$ \\
 B7 & IRIP & $5 \cdot 10^{-4}$ & 5 & $10^{-2}$ & 10855 & $5 \cdot 10^{-5}$ \\
 B8 & IRIP & $5 \cdot 10^{-4}$ & 5 & $10^{-2}$ & 10855 & $5 \cdot 10^{-4}$ \\
 B9 & IRIP & $5 \cdot 10^{-4}$ & 5 & $10^{-2}$ & 10855 & $5 \cdot 10^{-3}$ \\
 \hline
 C1 & IRIP & $1.5 \cdot 10^{-3}$ & 5 & $2 \cdot 10^{-3}$ & 9855 & $5 \cdot 10^{-5}$ \\
 C2 & IRIP & $1.5 \cdot 10^{-3}$ & 5 & $4 \cdot 10^{-2}$ & 11855 & $5 \cdot 10^{-5}$ \\
 C3 & IRIP & $1.5 \cdot 10^{-3}$ & 5 & $10^{-1}$ & 12855 & $5 \cdot 10^{-5}$
\end{tabular}
\end{ruledtabular}
\footnotetext[1]{These data are the effective values of the two-fluid parameters $\alpha_c$ and $\xi_p$ for the single-fluid cases, which are chosen as limits for the PIP simulations.}
\footnotetext[2]{$\xi_p$ (0) and $T_0$ are not independent variables. The initial ion fraction is determined by setting the value for $T_0$ at the beginning of the calculation.}
\end{table}

\subsection{Current sheet and reconnection rate}
\label{sec:reference_cases}

We analyse three PIP cases (listed as A1, A2 and A3 in Table \ref{tab:parameters_2D}) where different coupling terms are included in each simulation: A1 corresponds to a PIP case with elastic collisions only (NIR) and its parameters are set equal to the reference PIP cases examined in our previous work\cite{doi:10.1063/5.0032236}; A2 includes ionisation and recombination processes (IR); A3 includes the ionisation potential (IRIP). In the cases with ionisation and recombination processes (A2 and A3) we set $\tau_{_{\operatorname{IR}}} = 10^{-3}$. By choosing this value for $\tau_{_{\operatorname{IR}}}$ the ratio of the collision time to recombination time is $10^{-5}$, which is consistent with chromospheric rates\cite{Carlsson_2002, 2021A&A...645A..81S}. Based on the initial conditions, neutral-ion collisions occur on timescales of $\Delta t = (\alpha_c \rho_p)^{-1} = 1$, ion-neutral collisions on timescales of $\Delta t = (\alpha_c \rho_n)^{-1} \sim 0.01$ and recombination on time scales of $\Delta t = 10^3$.

The coalescence instability is shown in Figure \ref{fig:contour_Jz} at three different stages of development. In the first phase of the process the plasmoids come together, and when the oppositely directed magnetic field lines are pushed against each other a current sheet forms in the centre of the domain. In the first row of panels ($a$, $b$ and $c$) corresponding to the end of this first phase of coalescence, the current sheet appears as the thin red vertical region at $x = 0$. After the current sheet formation, reconnection begins and the current sheet slowly reduces in length, together with the decrease in size of the plasmoids. The reconnected field lines form an envelope that surrounds the two merging plasmoids and constitute the external boundary of the final plasmoid resulting from the coalescence. The second row of panels ($d$, $e$ and $f$) shows the plasmoid merging at these later stages, while the third row ($g$, $h$ and $i$) displays the end of coalescence and the larger merged plasmoid.

Each row of panels in Figure \ref{fig:contour_Jz} allows to compare the three PIP cases at similar stages of coalescence. Panels ($a$), ($b$) and ($c$) show the current sheet structure and early small-scale dynamics at the beginning of the reconnection phase. The current sheet thickness $\delta$ is estimated by taking the full width at 1/8 of the maximum current density $J_z$ along the $x$-axis ($y = 0$): this particular ratio is chosen to be consistent with the analysis previously performed on 2.5D calculations\citep{doi:10.1063/5.0032236}. A first difference between the three cases is observed in $\delta$ at the stage displayed by panels ($a$), ($b$) and ($c$). The current sheet thickess for case A1 is $\delta_{_{\operatorname{NIR}}} = 0.018$ ($\sim 9$ grid points), while $\delta_{_{\operatorname{IR}}} = 0.045$ ($\sim$ 23 grid points) and $\delta_{_{\operatorname{IRIP}}} = 0.029$ ($\sim$ 15 grid points) for cases A2 and A3, respectively. Case A1 displays the thinner current sheet and it is also the only case where the tearing instability develops. The inclusion of ionisation and recombination processes stabilises the current sheet against the tearing instability, as the plasma density increases in the reconnecting current sheet following a burst in ionisation (panels $b$ and $c$ of Figure \ref{fig:contour_Jz}). Varying the type of coupling between the fluids, from elastic collisions only to the inclusion of $\Gamma_{ion}$, $\Gamma_{rec}$ and ionisation potential, the coalescence dynamics also drastically changes. Formation and expulsion of large secondary plasmoids are observed in panel ($a$) of Figure \ref{fig:contour_Jz} for case A1. No plasmoids form in cases A2 and A3.

\begin{figure}[htb]
    \centering
    \includegraphics[width=\columnwidth,clip=true,trim=0cm 0cm 0cm 0cm]{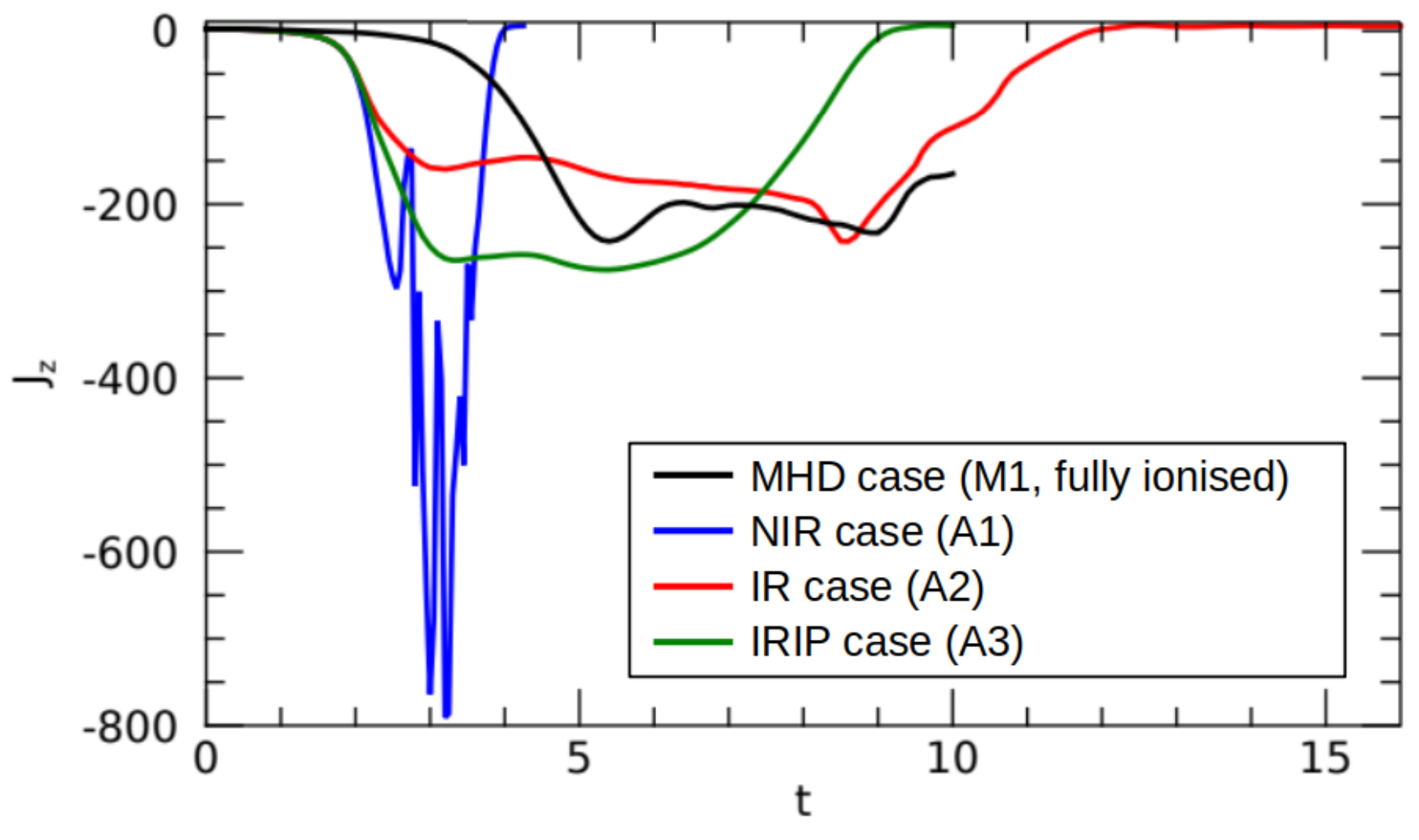}  
        \caption{Time evolution of current density $J_z$ at the centre of the current sheet for the PIP cases A1, A2 and A3 with $\alpha_c = 100$. The NIR case (blue), IR case (red) and IRIP (green) are compared to a MHD case (M1, black line), included as reference for the limit case of completely coupled fluids ($\alpha_c \rightarrow \infty$). The IR and IRIP cases are run at $\tau_{_{\operatorname{IR}}} = 10^{-3}$.}
    \label{fig:alpha_100_central_Jz}
\end{figure}

The coalescence time scale varies when varying the type of coupling between the fluids, as shown by panels ($g$), ($h$) and ($i$) where the final stage of the merging is displayed. Figure \ref{fig:alpha_100_central_Jz} shows the time variation of the current density at the centre of the current sheet for simulations A1, A2 and A3. These are compared to a single ﬂuid MHD case (listed as M1 in Table \ref{tab:parameters_2D}). M1 can be treated as the limit for a completely coupled system ($\alpha_c \rightarrow \infty$, $\tau_{_{\operatorname{IR}}} \rightarrow \infty$), where density and pressure are assumed to be equal to the bulk (ion + neutral) values.
The beginning of the reconnection phase is identified in all curves in Figure \ref{fig:alpha_100_central_Jz} by the first minimum in the current density. The formation of the central current sheet begins at similar times for all PIP simulations, suggesting that ionisation and recombination rates effects become relevant in the coalescence dynamics only once the current sheet is formed. During the initial phase of coalescence, the temperature is roughly constant and hence $\Gamma_{ion}$ and $\Gamma_{rec}$ do not vary much. During the second phase, where compressional, Ohmic and frictional heating become substantial, the temperature variation results in greatly enhanced ionisation and recombination rates (see Equations \ref{eq:recombination_rate} - \ref{eq:ionisation_rate}).

\begin{figure}[htb]
    \centering
    \includegraphics[width=\columnwidth,clip=true,trim=0cm 0cm 0cm 0cm]{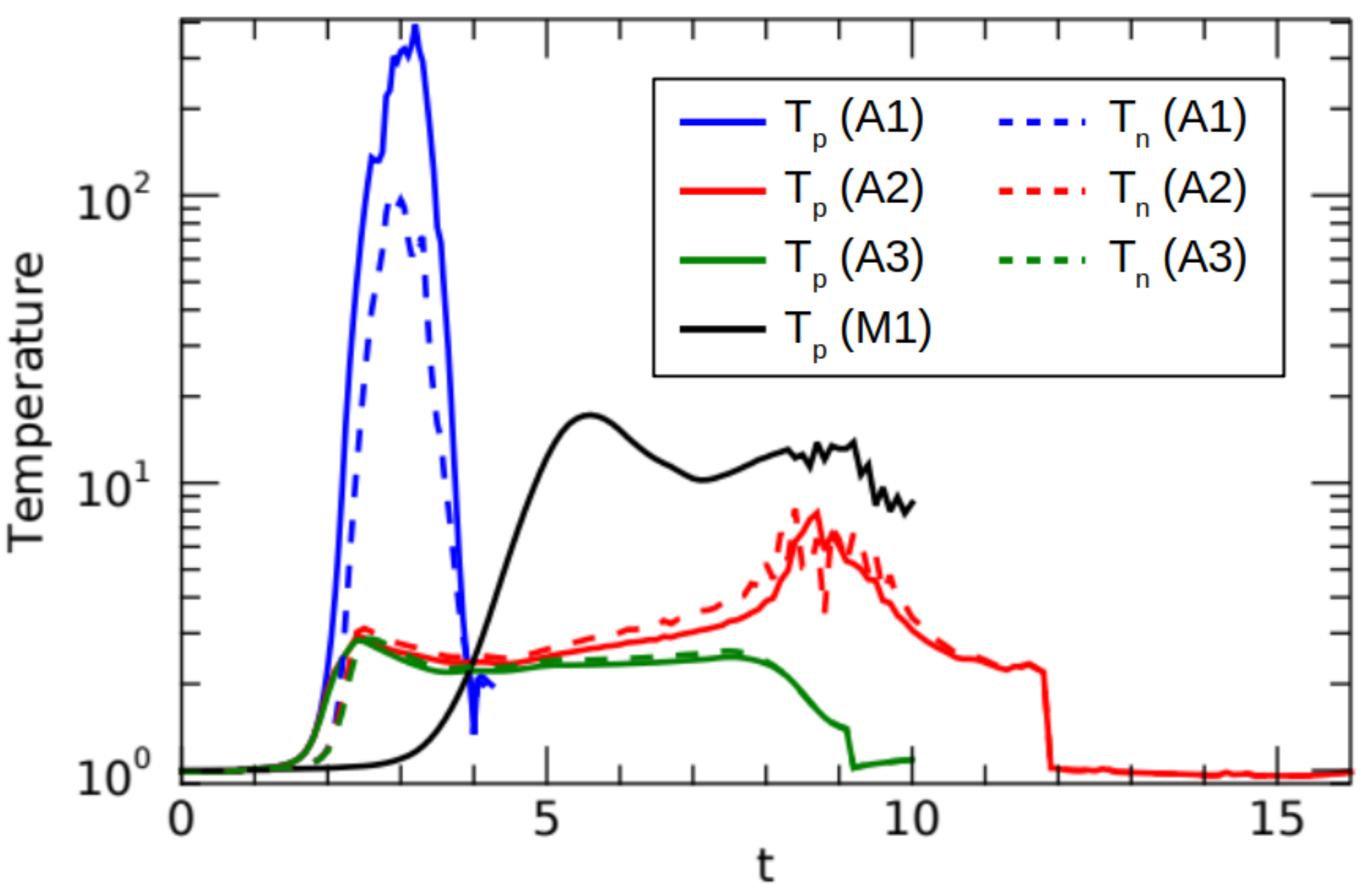}  
    \caption{Time variation of the mean plasma and neutral temperatures inside the current sheet for the PIP cases A1 (blue), A2 (red) and A3 (green), and the mean plasma temperature inside the current sheet of the MHD case M1 (black). Plasma temperatures are identified by the solid lines, neutral temperatures are identified by the dashed lines.}
    \label{fig:alpha_100_temperature}
\end{figure}
Figure \ref{fig:alpha_100_temperature} shows the mean plasma and neutral temperature in the current sheet of our fiducial MHD and PIP simulations as they vary with time. Initially, the temperature spikes in all simulations as the current sheet collapses, due to a combination of compressional and Ohmic heating, which dominate inside the current sheet, and frictional heating. The values for $T_p$ and $T_n$ in case A1 (NIR model) are consistent with the ones found in our previous study \citep{doi:10.1063/5.0032236}. The large increase in both plasma and neutral temperatures results from the small value of the ion plasma $\beta$ and the rarefaction of the reconnection region, where the density of both fluids is very low. The neutral rarefaction observed in case A1, consistent with the results shown in Figure 4 of our previous work\citep{doi:10.1063/5.0032236}, is the result of a strong divergence of the neutral velocity field. This divergence reaches values of $1.6$ inside the forming current sheet at $t = 2$ and increases to be $16.7$ at $t = 2.4$, and occurs because the neutrals are subject to two forces, generated by the drag from the plasma and by the neutral pressure respectively. As Figure 22 of our recent paper\citep{doi:10.1063/5.0032236} shows, the neutral pressure gradient works against the drag force to hinder the neutral inflow, but it works with the drag to promote the acceleration of the material in the outflow. As result, the neutrals are able to leave the current sheet faster than they enter, resulting in the current sheet becoming rarefied.

In cases A2 (IR model) and A3 (IRIP model) both $T_p$ and $T_n$ are much lower than case A1, and there is far less temperature difference between the two species. This happens because the strong heating in the current sheet has a large effect on the ionisation-recombination rates, which in turn act on the current sheet thickness: the high ionisation rate converts neutrals to plasma which results in the current sheet thinning less, $J_z$ is smaller than NIR cases (as shown in Figure \ref{fig:alpha_100_central_Jz}), and the Ohmic heating is smaller, thus leading to a cooler plasma.

In our equilibrium plasma at $t = 0$, the rates of both cases A2 and A3 are $\Gamma_{ion} = 10^{-5}$ and $\Gamma_{rec} = 10^{-3}$, and ionisation and recombination happen on time scales of $ \Delta t \sim 10^5$ and $10^3$ respectively; in our current sheet, ionisation happen on time scales of $\Delta t \sim$ 0.02 for case A2 ($\Gamma_{ion} = 47.12$) and $0.05$ for case A3 ($\Gamma_{ion} = 19.54$), while recombination occurs on time scales of $\Delta t \sim 8$ for both cases ($\Gamma_{rec} = 0.13$ in A2 and 0.12 in A3). This means that, for both cases A2 and A3, $\Gamma_{ion}$ varies of about 6 orders of magnitude while $\Gamma_{rec}$ varies of about 2 orders of magnitude from the initial phase to the reconnection phase. As the ionisation rate increases, so do the energy losses due to ionisation potential. The ionisation potential term acts to remove energy from the plasma, hence lowers $T_p$. In case A3, the cooling action of the ionisation potential contributes to further decrease both $T_p$ and $T_n$ compared to case A2. Including the physics of ionisation, recombination and ionisation potential leads to a more realistic description of the system, with a temperature variation of a factor of 3 rather than a factor of 400 as observed for the NIR model.

The end of the merging is identified for each case in Figure \ref{fig:alpha_100_central_Jz} by the current density acquiring a positive value again at later times, after the development of large negative currents during the reconnection phase. The fastest coalescence occurs in case A1 (blue curve in Figure \ref{fig:alpha_100_central_Jz}). The shorter time scale depends on the onset of turbulent reconnection, as the secondary plasmoids expelled from the current sheet allows a more efficient release of magnetic flux. Turbulent reconnection leads to a larger $J_z$ magnitude: the large fluctuations occurring at $t > 2.8$ are produced by secondary plasmoids forming and being ejected from the current sheet. In both case A2 (red) and A3 (green) reconnection occurs laminarly, similar to the MHD simulation. The coalescence time scale shortens at the inclusion of the ionisation potential. This is shown by both the comparison of the final time in panels (h) and (i) in Figure \ref{fig:contour_Jz} and the comparison between green and red curves in Figure \ref{fig:alpha_100_central_Jz}.  The shortening of the coalescence time scale occurs as the cooling of the ionisation potential results in the recombination of a large portion of ions and consequently a thinner current sheet.

\begin{figure}[htb]
    \centering
    \includegraphics[width=\columnwidth,clip=true,trim=0cm 0cm 0cm 0cm]{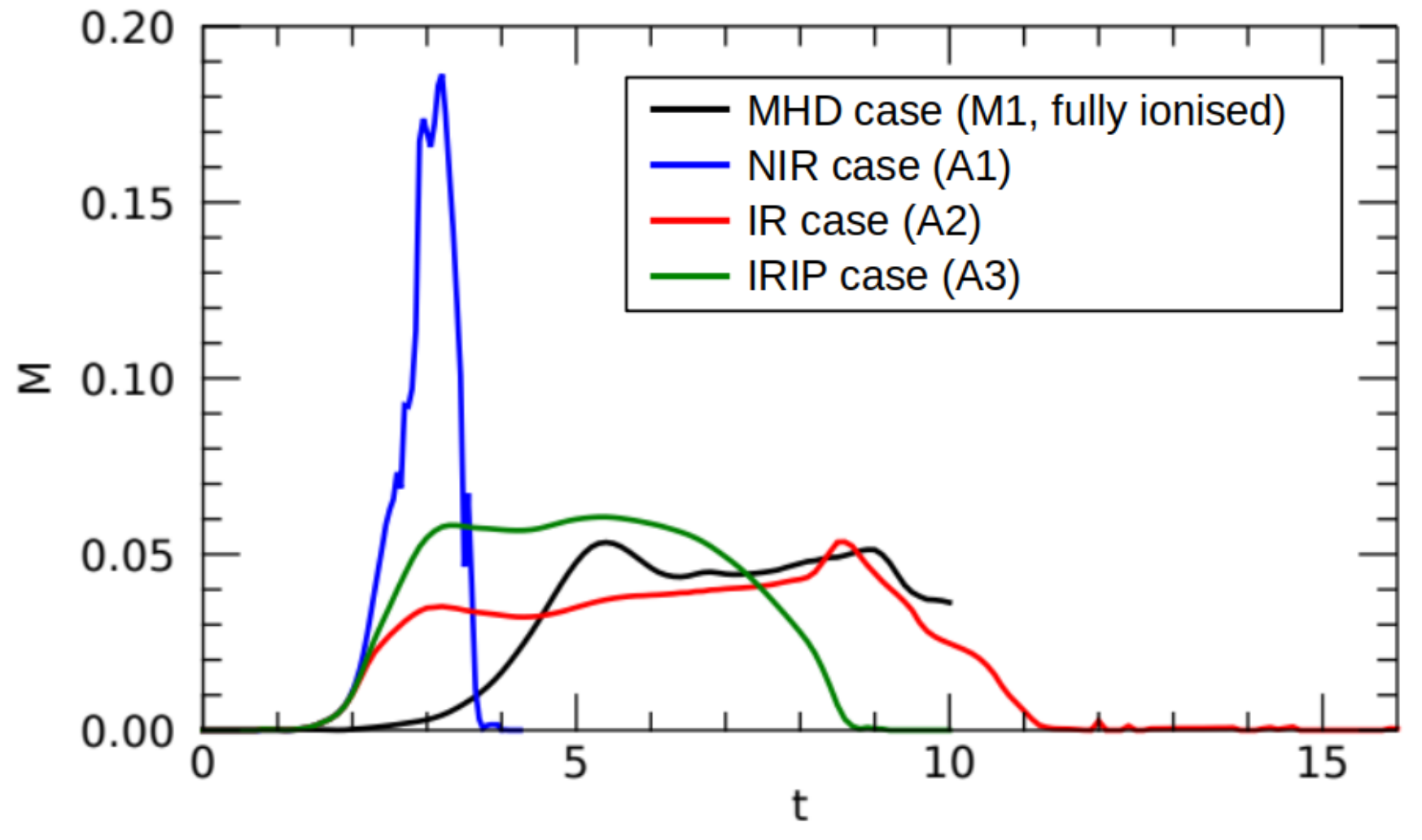}  
    \caption{Time variation of the reconnection rate for the PIP cases A1 (blue), A2 (red) and A3 (green) and the MHD case M1 (black).}
    \label{fig:alpha_100_recrate}
\end{figure}

Figure \ref{fig:alpha_100_recrate} shows the reconnection rate for simulations A1 (blue), A2 (red) and A3 (green) compared to the fully ionised plasma case M1 (black), where the reconnection rate as
\begin{equation}
    M = \frac{\eta J_{max}}{v_A B_{up}},
    \label{eq:reconnection_rate}
\end{equation}
where we chose $J_{max}$ to be the absolute maximum value of the current density inside the current sheet, $v_A$ the initial maximum bulk Alfv\'en speed and $B_{up}$ the initial maximum value of $B_y$ in the inﬂow.
Comparing the trend of the reconnection rate in Figure \ref{fig:alpha_100_recrate} with the evolution of the current density previously shown in Figure \ref{fig:alpha_100_central_Jz}, it is evident that for case A1 the reconnection rate shows far less fluctuation than $J_z$. This comes from the definition of the reconnection rate, which depends on $J_{max}$: the maximum current density inside the current sheet is not necessarily occurring at its centre at all times, and this is particularly evident in case of secondary plasmoids production. While we had fixed the calculation of $J_z$ at the centre to evaluate the type of reconnection developing in the current sheet, we now want to examine the maximum reconnection rate that is achieved within the current sheet.

The mean reconnection rate of case A1, where the fluids are coupled through elastic collisions and charge exchange only, is approximately $M_{_{\operatorname{NIR}}} = 0.12 \pm 0.05$: in this case the reconnection rate displays very sharp variations during the merging following the formation and ejection of the secondary plasmoids. The stabilisation of the current sheet by the action of ionisation and recombination is reflected in the lower peak value and flatter trend of the reconnection rate for the cases A2 and A3, where secondary plasmoids do not form and whose smoother fluctuations can be compared to the MHD case. The mean reconnection rates of cases A2 and A3 are also comparable to the MHD rate, $M_{_{\operatorname{MHD}}} = 0.047 \pm 0.003$. In case A2, $M_{_{\operatorname{IR}}} = 0.038 \pm 0.006$, lower than the fully ionised case. As already shown by the current density in Figures \ref{fig:contour_Jz} and \ref{fig:alpha_100_central_Jz}, coalescence in case A2 occurs over a longer time than case A1. The longer coalescence time scale is reflected in the reconnection rate, which steadily increases with time and does not show violent fluctuations, as shown by the red curve in Figure \ref{fig:alpha_100_recrate}. The addition of the ionisation potential in case A3 speeds up the coalescence with respect to case A2, as the current sheet thins under the action of cooling and recombination, and is more compressed by the inflow. The higher mean reconnection rate of case A3, $M_{_{\operatorname{IRIP}}} = 0.057 \pm 0.003$, is consistent with the shorter coalescence time scale observed in Figure \ref{fig:alpha_100_central_Jz}.

In this Section we have extensively discussed the differences between NIR, IR and IRIP coupling models for partially ionised plasmas during plasmoid coalescence. To summarise our key results, we find that ionisation and recombination have a stabilising effect on the current sheet, which is not often included in the linear theory description. Compared to fully ionised plasmas at the same bulk density, ionisation and recombination still lead to a faster reconnection in PIP, and the main effects are observed in the second phase of coalescence (reconnection phase).

\subsection{Oscillatory behaviours of the current sheet}

Several types of oscillatory motions develop during plasmoid coalescence. Waves are produced in all simulations, and are particularly evident in case A2 (IR model, intermediate initial coupling $\alpha_c = 100$), as shown by panels ($e$) and ($h$) of Figure \ref{fig:contour_Jz}. Smaller scale oscillations occur in the reconnection region and along the current sheet. In this Section we describe the two main oscillatory motions that develop in the current sheet.

\begin{figure}[htb]
    \centering
    \includegraphics[width=\columnwidth,clip=true,trim=0cm 0cm 0cm 0cm]{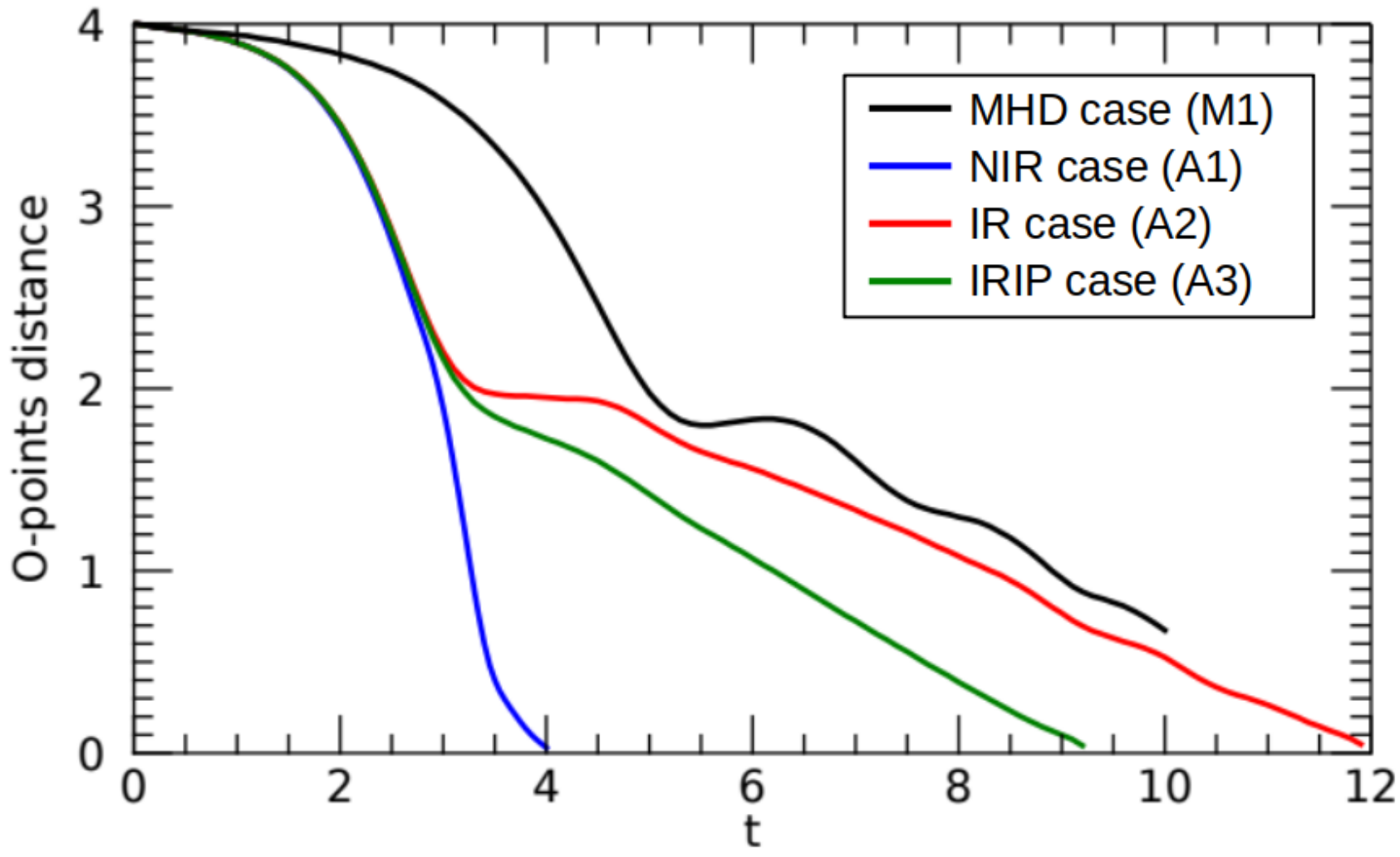}  
    \caption{Time variation of the distance between the merging plasmoids, calculated as the distance between the $O$-points for the MHD case (black) and the PIP cases A1 (blue), A2 (red) and A3 (green).}
    \label{fig:O_point_separation}
\end{figure}

A first type of oscillations is linked to the increase of gas pressure in the current sheet due to the plasmoids moving closer. In our previous work we observed a regular fluctuation in the $O$-points separation for the MHD case\citep{doi:10.1063/5.0032236}, while such motion was suppressed in the PIP case due to the faster reconnection rate. When ionisation and recombination are included (A2 and A3 cases) we find an intermediate situation between the regular peaks as in MHD and their complete absence in NIR cases, with an initial large fluctuation shown in Figure \ref{fig:O_point_separation} and smaller oscillations that are damped more quickly than the MHD case. The first oscillation corresponds to the large local maximum observed at the beginning of the reconnection phase for case A2 and A3, and it is a direct consequence of the processes involved in the current sheet formation. In fact, during the current sheet formation the plasma temperature increases in the thin region between the two initial plasmoids: the higher temperature leads to an initial burst in ionisation that increase the plasma pressure at the centre of the current sheet. When the plasma pressure becomes sufficiently high, it balances the pressure of the plasmoids moving closer and the plasmoids rebound off each other.
At later stages, irregular oscillations of smaller magnitude can still be observed in the IR case (A2, red curve in Figure \ref{fig:O_point_separation}), while in the IRIP case (A3, green curve) the $O$-points distance decreases steadily.

\begin{figure}[htb]
    \centering
    \includegraphics[width=\columnwidth,clip=true,trim=0cm 0cm 0cm 3cm]{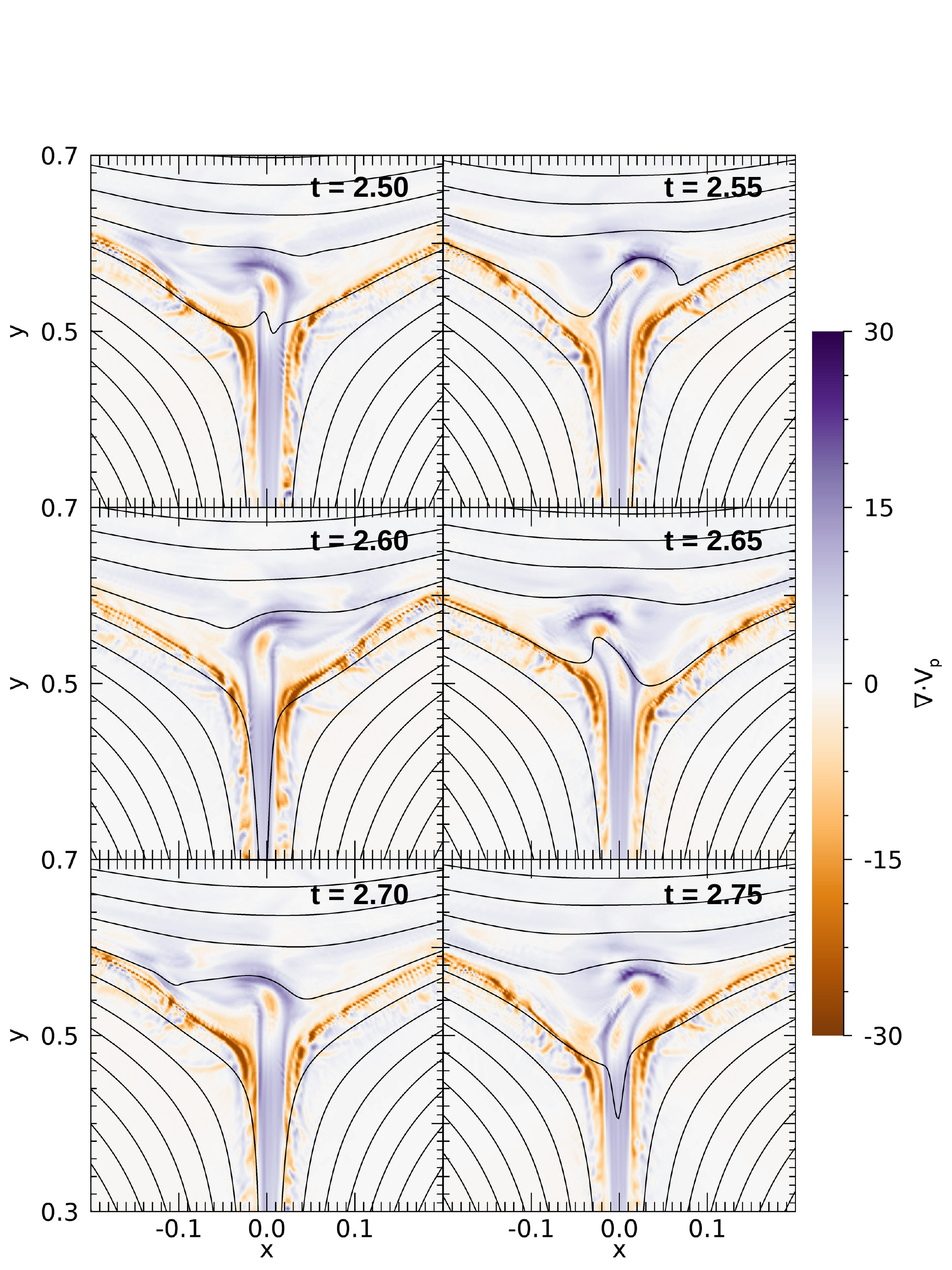}  
    \caption{Time frames of the plasma velocity divergence shown at the current sheet upper edge for the IRIP case B2. Magnetic field lines are shown in black.}
    \label{fig:vel_div}
\end{figure}

In the reconnection phase of the majority of our simulations, a second type of oscillations is observed at later times during the merging. This type of motion is displayed in Figure \ref{fig:vel_div}, where the divergence of the plasma velocity is displayed in the upper half of the current sheet for a reference case (B2).
Oscillations start at the jet termination shock, rapidly increasing in amplitude, and the motion propagates towards the centre of the current sheet. The displacement of the current sheet centre along the $x$-axis can be seen at $y =0.3$ in all the frames of Figure \ref{fig:vel_div}. For simplicity we have displayed oscillations in a small part of the domain around the current sheet. However, analogue oscillations are observed in the bottom half of the current sheet.

The oscillations are spatially resolved. We measure the displacement of the current sheet along $x$ due to the oscillations by tracking the position in time of the peak in $|J_z|$, which corresponds to the central point of the current sheet width. In case B2, half an oscillation occurs in about $\Delta t = 0.1$, as suggested by Figure \ref{fig:vel_div}. At $y = 0.3$ the distance between the peak at $t = 2.50$ (first panel in Figure \ref{fig:vel_div}) and the peak at $t = 2.60$ (third panel in Figure \ref{fig:vel_div}) is $\Delta l = 7.8 \cdot 10^{-3}$, which is four times bigger than the grid size $\Delta x = 1.95 \cdot 10^{-3}$. Moving towards the end of the current sheet, the oscillation amplitude increases, therefore oscillations are resolved by a larger number of grid points.

In cases with very thin current sheets, such as A1 (NIR model) and the IRIP cases B1, B7, B8 and B9 in Table \ref{tab:parameters_2D}, turbulent reconnection takes place before oscillations can develop. In fully ionised plasmas, where the current sheet is thicker, the oscillatory motion can be suppressed by increasing $\eta$. We display this change of regime by running an MHD case (M2) where we increase $\eta$ by a factor of 2 ($\eta = 0.003$) from the value set for case M1 (see Table \ref{tab:parameters_2D}). The current sheet of case M2 is subject to a steady, laminar reconnection, where no oscillations develop. The oscillatory behaviour is therefore constrained by the diffusivity, whose value determines the evolution of the current sheet dynamics into either a turbulent process (lower $\eta$) or steady reconnection (higher $\eta$).

Recent high-resolution MHD simulations of fully ionised plasmas \cite{2017ApJ...848..102T} have found localised nonlinear oscillations at the edges of current sheets, studied in the framework of the dynamics of flux rope eruption in solar atmospheric plasmas. In this recent study\cite{2017ApJ...848..102T} laminar reconnection occurs for a global Lundquist number $S = 2.8 \cdot 10^3$ and oscillations develop when the global $S = 5.5 \cdot 10^3$, while at high values of $S$ ($S = 2.8 \cdot 10^4$) plasmoids form in the reconnecting current sheet \cite{2017ApJ...848..102T}. The Lundquist numbers of our simulations are calculated by using the effective Alfv\'en speed $v_{A,e} \sim v_{\operatorname{out}}$, where $v_{\operatorname{out}}$ is the ion outflow speed. This particular choice for approximating the Alfv\'en speed, consistent with our previous work\citep{doi:10.1063/5.0032236}, accounts for the effective density given by the partial coupling between plasma and neutrals in the PIP cases. In our configuration we observe an oscillatory behaviour for Lundquist numbers between $2.6 \cdot 10^3$ and $3.3 \cdot 10^3$ in partially ionised plasma cases, while the MHD case develops this dynamics at $S = 3 \cdot 10^3$. 

\begin{figure}[htb]
    \centering
    \includegraphics[width=\columnwidth,clip=true,trim=0cm 7.5cm 0cm 8.5cm]{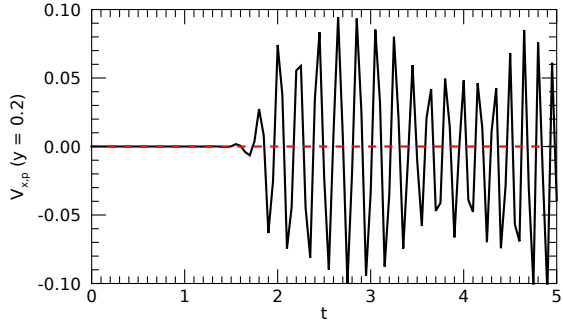}
    \caption{Time variation of $v_{x,p}$ at $x = 0, y = 0.2$ for the IRIP case B2. This indicates the displacement of the current sheet vertical axis in the $x$-direction. The red dashed line indicates the value of $v_{x,p} = 0$.}
    \label{fig:oscillations_plot}
\end{figure}

In order to better characterise the oscillatory motion, we look at the shift of the current sheet vertical axis position along $x$ at a small distance from its centre. Figure \ref{fig:oscillations_plot} shows the time variation of the plasma $v_x$ at $x = 0$ and $y = 0.2$, closer to the centre of the current sheet than the region shown in Figure \ref{fig:vel_div}.
As no grid point lays exactly at $x = 0$, but two sets of grid points are located at a symmetric distance from the coordinates origin, we perform a linear interpolation of the grid points located symmetrically across the $y$-axis. The interpolation allows to cancel out the $v_{x,p}$ contributions at the same magnitude but opposite in sign at the centre of the current sheet and identify the real oscillations that displace the upper part of the structure.

The beginning of the oscillatory motion is identified in this work with the position in time of the first peak having an amplitude $v_{x,p} > 0.001$. We chose this threshold value as it is twice the magnitude of the initial white noise perturbation, so that random motions along the $x$-axis would not be mistaken with the beginning of the oscillations. In the case of the simulation B2 oscillations begin at $t = 1.55$.

We calculate the period of oscillations as the mean distance between the peaks of $v_{x,p}$ shown in Figure \ref{fig:oscillations_plot}. The oscillation periods and their error, calculated as the standard deviation of the peak separation sets of measures, are reported in Table \ref{tab:parameters_oscillations}. Looking at the cases in the initial survey presented in Section \ref{sec:reference_cases}, case A2 shows a longer period, $P_{A2} = 0.39 \pm 0.07$, than case A3, where the period is $P_{A3} = 0.22 \pm 0.04$. A period of $\sim 0.2$ has been found for all the IRIP simulations run at $\alpha_c = 5$, as shown in Table \ref{tab:parameters_oscillations}, which is consistent with case A3 run at $\alpha_c = 100$.
\begin{table}
\caption{Parameters of the oscillatory behaviour for the 2.5D simulations. The oscillation period is presented with its standard deviation \label{tab:parameters_oscillations}}
\begin{ruledtabular}
\begin{tabular}{ccccc}
 ID & $t_{\operatorname{peak}}$ & $P$ & $\delta (t = t_{\operatorname{ peak}}) $ & $L (t = t_{\operatorname{ peak}}) $ \\
\hline
 M1 & 7.1 & 0.44 $\pm$ 0.09 & 0.041 & 0.78 \\ 
 \hline
 A2 & 5.8 & 0.39 $\pm$ 0.07 & 0.043 & 1.21 \\
 A3 & 5.7 & 0.22 $\pm$ 0.04 & 0.029 & 1.07 \\
 \hline
 B2 & 1.55 & 0.19 $\pm$ 0.03 & 0.029 & 1.30 \\
 B3 & 2.10 & 0.20 $\pm$ 0.03 & 0.029 & 1.26 \\
 B4 & 1.90 & 0.20 $\pm$ 0.03 & 0.025 & 1.28 \\
 B5 & 1.75 & 0.15 $\pm$ 0.03 & 0.021 & 1.29 \\
 B6 & 1.25 & 0.17 $\pm$ 0.09 & 0.021 & 1.28 \\
 \hline
 C1 & 1.15 & 0.20 $\pm$ 0.03 & 0.027 & 1.30 \\
 C2 & 2.35 & 0.17 $\pm$ 0.04 & 0.025 & 1.24 \\
 C3 & 3.35 & 0.14 $\pm$ 0.03 & 0.025 & 0.95
\end{tabular}
\end{ruledtabular}
\end{table}
In the MHD case, the oscillation period is $P_{_{\operatorname{MHD}}} \sim 0.44$, which is comparable with the IR case A2, but approximately double than all of the IRIP cases. All the simulations are highly temporally resolved, however the cadence of the time output is longer than the time step, with an output being saved approximately every $10^3 - 10^4$ iterations. As our cases are analysed post process, the chosen output might set limitations on the analysis of the period. The time outputs used in this work are $\Delta t = 0.1$ for cases M1, A2 and A3, and $\Delta t = 0.05$ for the remaining IRIP cases. These result in having about four points to determine a period in all the cases with the exception of case A3, where only two points define the period. However, the period length was confirmed through tests run by saving smaller time outputs, which prove that the oscillatory behaviour has the same period.

It can be suggested that the longer periods identified in both MHD and IR cases are dependent on current sheet properties. The current sheets length $L$ and thickness $\delta$ are reported in Table \ref{tab:parameters_oscillations} at the time $t_{\operatorname{peak}}$ of the first oscillation for all the cases. A correlation can be observed between oscillation period and $\delta$: the current sheet is thicker in M1 and A2, where the period is also longer, than the IRIP cases, where the cooling from the ionisation potential results in thinner current sheets.

\begin{figure}[htb]
    \centering
    \includegraphics[width=\columnwidth,clip=true,trim=0cm 0cm 0cm 0cm]{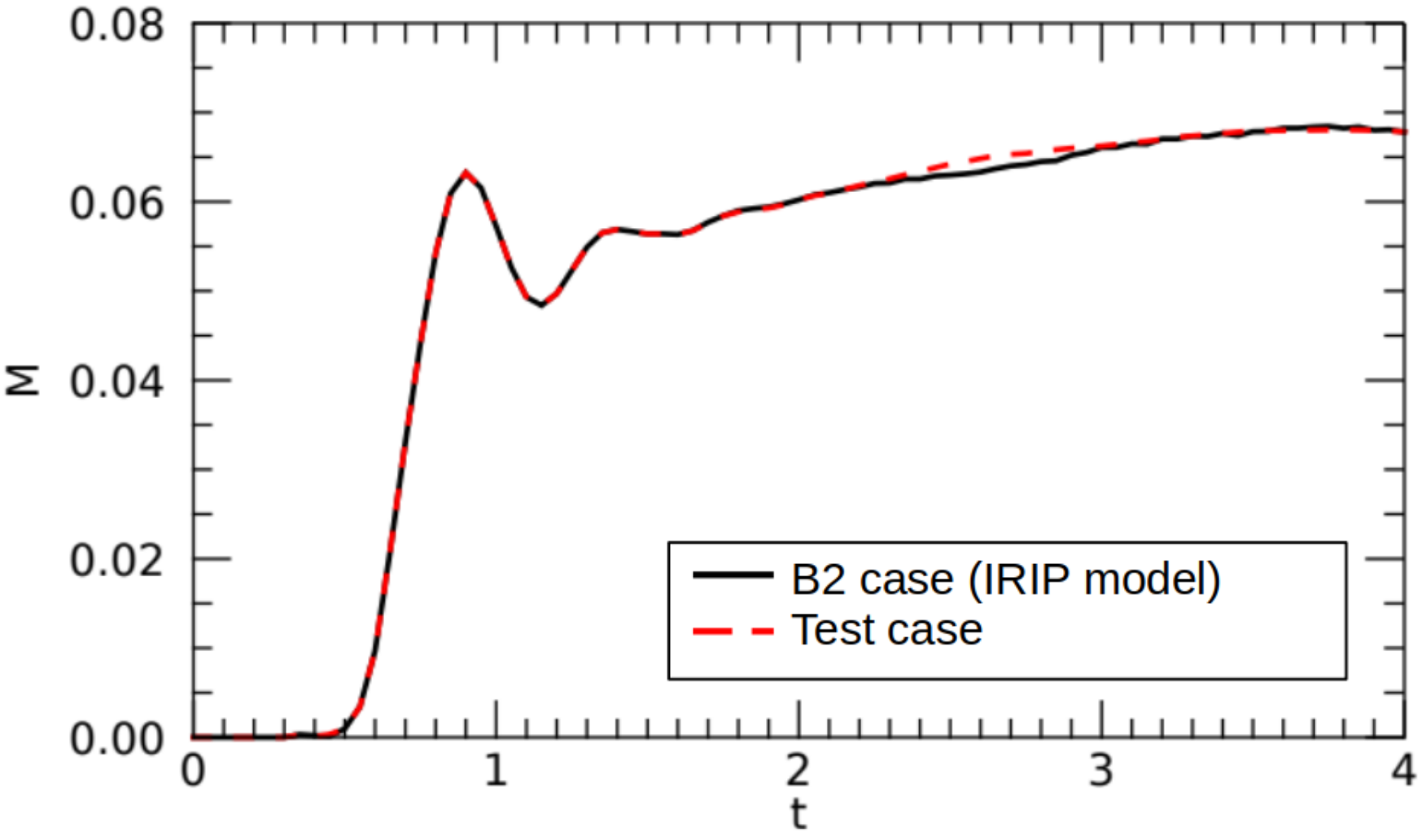}  
    \caption{Comparison of the time variation of the reconnection rate between the IRIP case B2 (solid black curve) and a test simulation run at half the domain of B2 (dashed red line). Oscillations in the test case are suppressed by the symmetric boundary set at $x = 0$.}
    \label{fig:oscillations_rate_comparison}
\end{figure}
The oscillatory dynamics does not have any influence on the reconnection rate. This can be seen through the comparison of case B2 with a second test simulation run with the same set of parameters and resolution. This second case is run in half domain in the $x$ direction ($x = [0, 4]$), with a symmetric boundary located at $x = 0$ that prevents the onset of oscillations. Figure \ref{fig:oscillations_rate_comparison} shows the reconnection rate $M$ of case B2 (solid black line) and the test simulation at half domain (dashed red line) where the symmetry suppresses the oscillatory motion. Both simulations display the same reconnection rate magnitude and variation across the same time interval, with minor fluctuations occurring at $t > 2$.

In this Section we have discussed the onset of oscillatory motions propagating in the current sheet. Previous works on fully ionised solar atmospheric plasmas \cite{2016ApJ...823..150T,2017ApJ...848..102T} had observed the onset of localised oscillations in reconnecting current sheets. We find that the oscillations developing at the edges of current sheets are independent of ionisation and recombination processes and more generally of partial ionisation. The oscillation period shows a direct correlation with the current sheet thickness, independently of the type of plasma chosen for the simulation.

\subsection{Secondary plasmoids}

\begin{figure}[htb]
    \centering
    \includegraphics[width=\columnwidth,clip=true,trim=1cm 0.5cm 0.9cm 0.8cm]{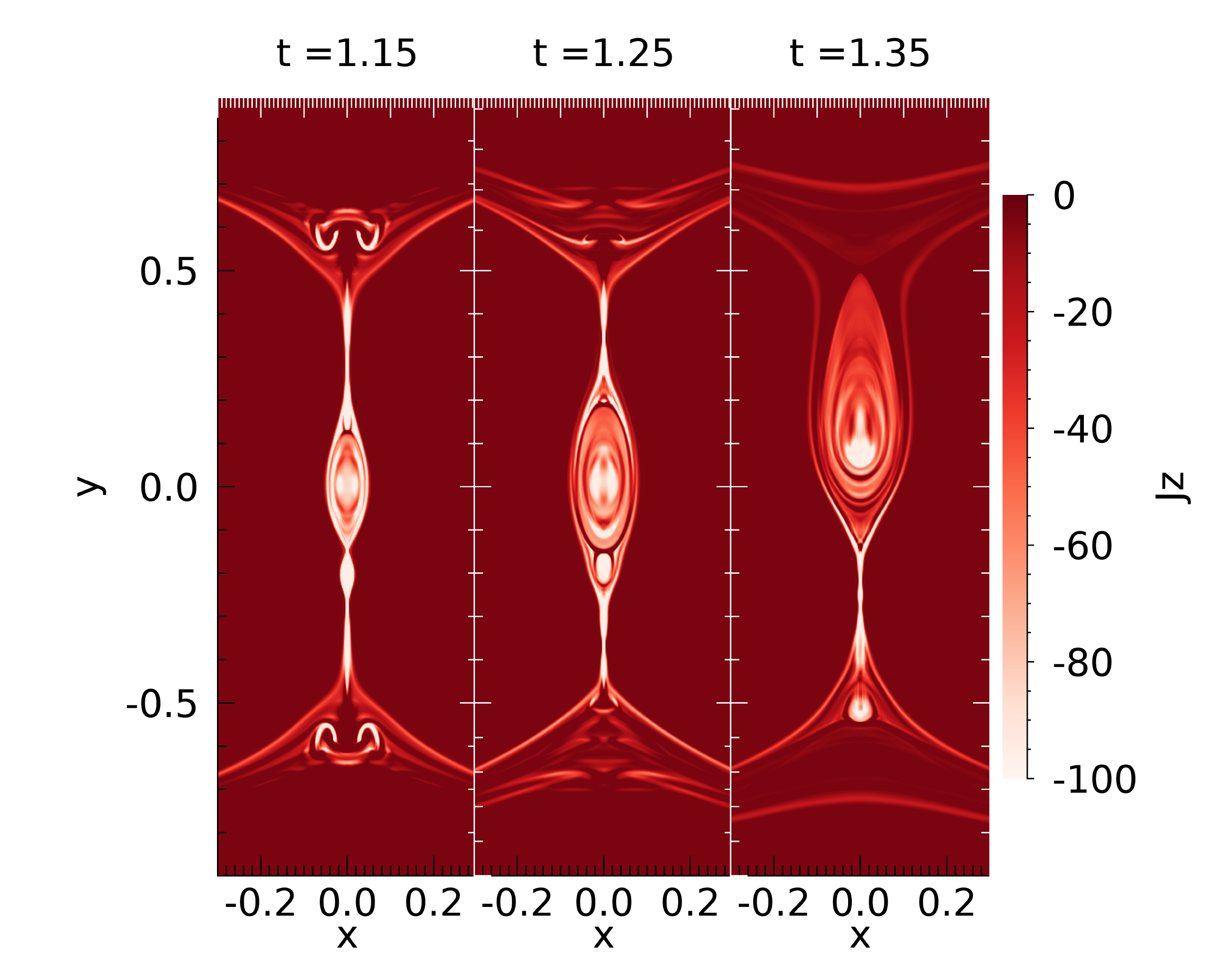}  
    \caption{Secondary plasmoid coalescence occurring in the central current sheet  for the IRIP case B7. Two plasmoids are present at $t = 1.15$ (left panel). At $t = 1.25$ (central panel) the plasmoids start merging, as reconnection takes place in the current sheet located in between them. Coalescence is completed at $t = 1.35$ (right panel), and the final plasmoid moves along the current sheet to be ejected.}
    \label{fig:secondary_plasmoids_B7}
\end{figure}

Evidence of fractal coalescence is observed in the IRIP cases B1, B7, B8 and B9, where the central current sheet is subject to the tearing instability and secondary plasmoids are produced. Figure \ref{fig:secondary_plasmoids_B7} shows the interaction of two secondary plasmoids coalescing before leaving the current sheet for case B7 in Table \ref{tab:parameters_2D}.

Given the overall effects of ionisation and recombination on plasmoid coalescence discussed in Section \ref{sec:reference_cases}, it is interesting to examine the secondary plasmoids and compare their properties to the ones observed in NIR simulations of our previous study \citep{doi:10.1063/5.0032236}. Ionisation and recombination provide additional force terms that can be analysed to investigate the equilibrium of secondary plasmoids. Therefore, we look at the force balance and the magnitude of $\Gamma_{ion}$ and $\Gamma_{rec}$ across the two secondary plasmoids appearing on the left panel of Figure \ref{fig:secondary_plasmoids_B7} ($t = 1.15$) and see if the new terms shift the force balance when compared to NIR cases \citep{doi:10.1063/5.0032236}.

\begin{figure}[htb]
    \centering
    \includegraphics[width=\columnwidth,clip=true,trim=2cm 3.5cm 1.2cm 2.5cm]{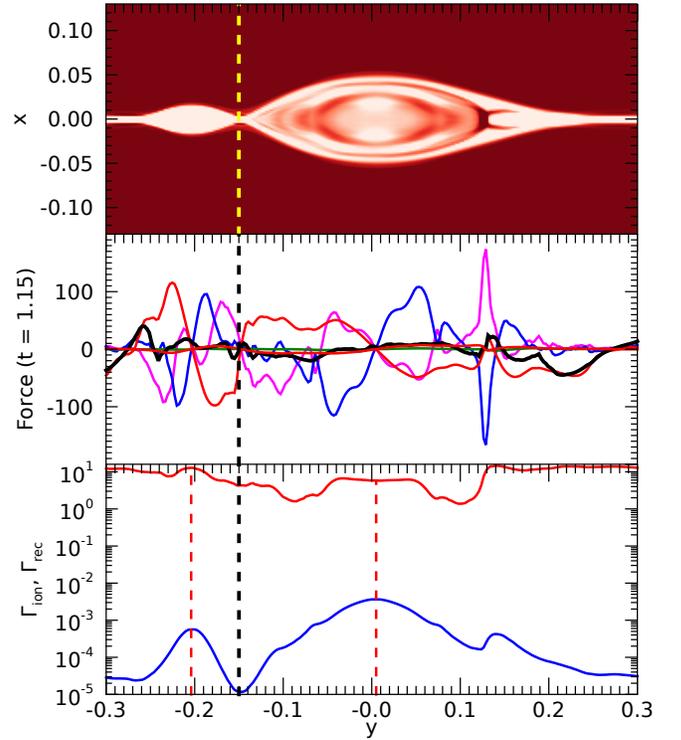}
    \caption{Top panel: detail of the current density magnitude of secondary plasmoids at t = 1.15 in the IRIP case B7. Central panel: force balance $\mathbf{J} \times \mathbf{B} - \nabla p$ (black solid line) calculated along the current sheet in the $y$-axis, compared to the position of secondary plasmoids in the top panel.  The force components are $- \nabla p_p$ (blue), $- \nabla p_n$ (green), magnetic pressure (magenta) and magnetic tension (red). Bottom panel: ionisation (red) and recombination (blue) rates along the current sheet, compared to the position of secondary plasmoids in the top panel. The yellow (top panel) and black (central and bottom panel) dashed line indicate the edge between the two plasmoids. The red vertical dashed lines in the bottom panel indicate the position of the secondary plasmoids centres along the $y$-axis.}
    \label{fig:force_balance}
\end{figure}

The force contributions and balance between the total pressure gradient and the Lorentz force ($\mathbf{J} \times \mathbf{B} - \nabla p$) are shown for the IRIP case B7 at $x = 0$ for $ t = 1.15$ in the central panel of Figure \ref{fig:force_balance}, compared to the current density magnitude map (top panel). The yellow and black vertical dashed lines in all panels are representative of the $X$-point location between the two plasmoids.
The force components cancel each other at the plasmoids location, while the current sheet around is still out of balance. Inside the plasmoids, the major contributions to the total force are provided by the gradient of the plasma pressure (blue curve), the magnetic pressure $B^2 / 2$ (magenta curve) and the $y$-component of the magnetic tension $(\mathbf{B} \cdot \nabla) \cdot \mathbf{B}$ (red curve). This situation is similar to the cases examined in our previous study\citep{doi:10.1063/5.0032236}: the outer structure is characterised by an almost force-free equilibrium, while $- \nabla p_p$ is significant around the plasmoid centre. However, a difference is observed in the distribution of the force components in the inner structure. At the plasmoid centre we observe a magnetohydrostatic equilibrium, with both magnetic pressure and magnetic tension balancing the plasma pressure gradient, while a force-free magnetic equilibrium is sustained at the edge, where magnetic pressure and magnetic tension balance each other. Comparing the observed structure with our previous results, we see that the core region is larger in the IRIP cases than in the NIR cases. In the IRIP case the force-free equilibrium occurs in an external thin annulus, while in NIR cases force-free equilibrium nearly entirely dominates the plasmoid structure, with the exception of a very small region at the plasmoid centre. This feature is especially evident in the larger plasmoid on the right.

$\Gamma_{ion}$ (red) and $\Gamma_{rec}$ (blue) are shown in the bottom panel of Figure \ref{fig:force_balance}. The centre of the two plasmoids, located at $y = -0.204$ for the smaller plasmoid and $y = 0.005$ for the bigger plasmoid respectively, are indicated by the two red vertical dashed lines. Recombination is observed at very small rates along the current sheet and at the centre of secondary plasmoids, while larger fluctuations are detected in the ionisation rate. Both $\Gamma_{ion}$ and $\Gamma_{rec}$ are larger in the central part of the plasmoids, where the inner structure is in magnetohydrostatic equilibrium, while the external ring characterised by a force-free magnetic equilibrium is subject to lower change rates. While $\Gamma_{rec}$ is larger at the plasmoid centre and decreases towards the ends of the inner structure, the magnitude of $\Gamma_{ion}$ in the bigger plasmoids tends to be slightly larger at the interface between inner and outer plasmoid region than in the very centre of the plasmoid.

\begin{figure}[htb]
    \centering
    \includegraphics[width=\columnwidth,clip=true,trim=1.5cm 8.5cm 1.8cm 8.7cm]{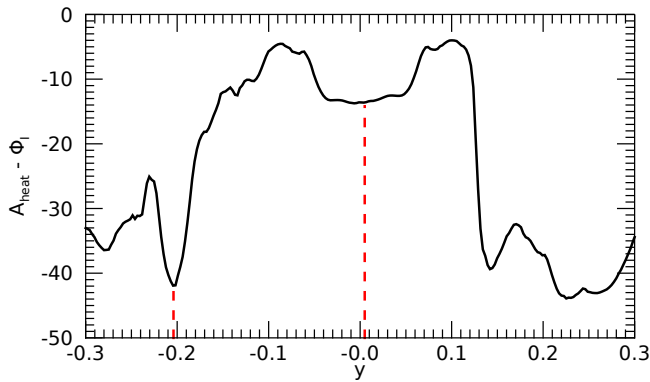}  
    \caption{Energy loss rate $A_{heat} -  \Phi_I$ calculated along the current sheet ($x = 0$) at $t = 1.15$ for the IRIP case B7. The red vertical dashed lines indicate the centre of the two secondary plasmoids interacting inside the current sheet.}
    \label{fig:energy_loss}
\end{figure}
Figure \ref{fig:energy_loss} shows the energy loss rate calculated as $A_{heat} -  \Phi_I$, defined by Equations (\ref{eq:Phi_I})-(\ref{eq:A_heat}), across the secondary plasmoids at $x = 0$ and $t = 1.15$. The cooling from the term $A_{heat} -  \Phi_I$ is stronger across the secondary plasmoids than the current sheet, as shown from the troughs in the energy loss rate around the location of the plasmoids centres (identified by the red dashed lines in Figure \ref{fig:energy_loss}). The action of the ionisation potential, which corresponds to a neat temperature decrease, can contribute to the increase in the recombination rate observed inside secondary plasmoids (see bottom panel of Figure \ref{fig:force_balance}).

We have seen that $\Gamma_{ion}$, $\Gamma_{rec}$ and the ionisation potential modify the force distribution inside secondary plasmoids compared to NIR cases \citep{doi:10.1063/5.0032236}. We want to evaluate whether the variation in these rates change the force balance inside secondary plasmoids enough to modify their structure and interaction before they leave the current sheet. From the energy loss rate we can estimate the cooling time of the plasmoid, which is calculated by dividing the total internal energy by the term $A_{heat} -  \Phi_I$. At the centre of the larger plasmoid, the cooling time is estimated to be $t \sim 0.7$. The expulsion time of the larger plasmoid from the current sheet is estimated to be $t \sim 0.2$, a third of the cooling time for the same plasmoid. Other smaller plasmoids coalesce or are expelled in shorter times. Therefore, we consider the secondary plasmoids to be approximately in equilibrium during their interaction inside the current sheet. Since the cooling time is larger than the expulsion time, ionisation-recombination effects do not really influence the force equilibrium of the plasmoid internal structure. Despite having a slightly different internal distribution, these secondary plasmoids act in the same way as in the NIR cases discussed in our previous study \citep{doi:10.1063/5.0032236}.

In summary, secondary plasmoids can form in the IRIP model when turbulent reconnection develops. The force balance inside these plasmoids is slightly changed under the action of ionisation, recombination and radiative losses. However, the plasmoids equilibrium and interaction inside the current sheet are unchanged when compared to previously investigated NIR cases \citep{doi:10.1063/5.0032236}. More details on the onset of turbulent reconnection in IRIP simulations are discussed in Section \ref{sec:tau_IR}.

\subsection{Survey on $\tau_{_{\operatorname{IR}}}$}
\label{sec:tau_IR}

The effects of the relative importance of the ionisation and recombination rates and the collision rates are investigated through a survey on the parameter $\tau_{_{\operatorname{IR}}}$, that is varied in the interval $[5 \cdot 10^{-6}, 5 \cdot 10^{-1}]$. The simulations listed in Table \ref{tab:parameters_2D} with IDs B1 to B6, characterised by a diffusivity $\eta = 1.5 \cdot 10^{-3}$, are compared to simulations run at a lower $\eta = 5 \cdot 10^{-4}$ that are listed with IDs B7 to B9. Preliminary tests on 1D current sheets had shown that at lower collisional coupling secondary plasmoids might still form, despite the stabilisation of ionisation and recombination on reconnection. We choose to study the coalescence for $\alpha_c = 5$, lower than the cases presented in Section \ref{sec:reference_cases}, as preliminary tests have identified this collisional coupling regime to better promote the onset of nonlinear dynamics in the IRIP cases.

\begin{figure}[htb]
    \centering
    \includegraphics[width=\columnwidth,clip=true,trim=0cm 0cm 0cm 0cm]{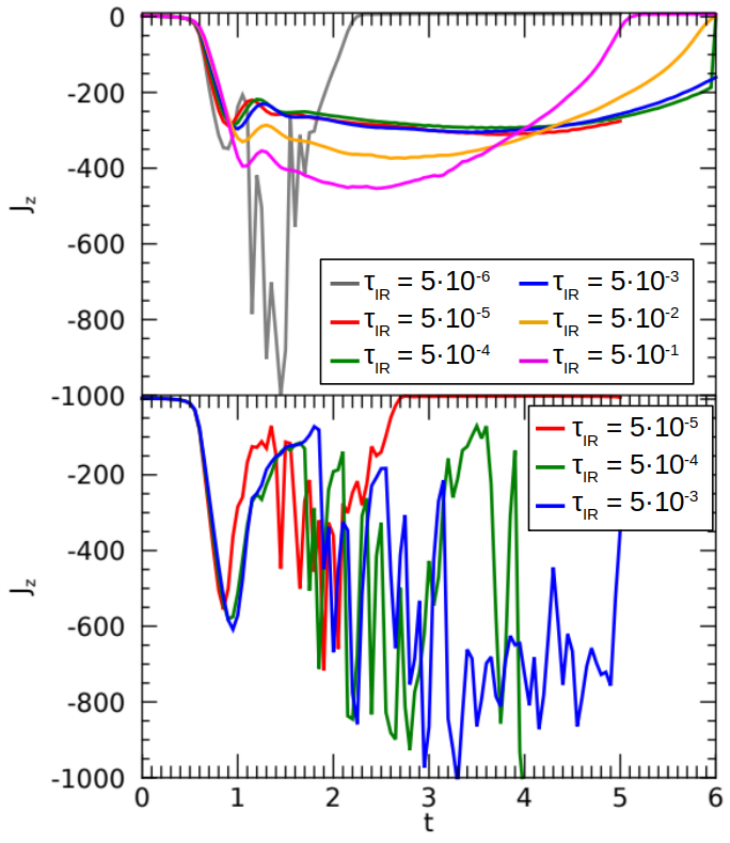}  
    \caption{Top panel: time evolution of $J_z$ at the centre of the current sheet for cases B1 to B6, run at $\eta = 1.5 \cdot 10^{-3}$. The simulations are run respectively at $\tau_{_{\operatorname{IR}}} = 5 \cdot 10^{-6}$ (gray), $\tau_{_{\operatorname{IR}}} = 5 \cdot 10^{-5}$ (red), $\tau_{_{\operatorname{IR}}} = 5 \cdot 10^{-4}$ (green), $\tau_{_{\operatorname{IR}}} = 5 \cdot 10^{-3}$ (blue), $\tau_{_{\operatorname{IR}}} = 5 \cdot 10^{-2}$ (orange) and $\tau_{_{\operatorname{IR}}} = 5 \cdot 10^{-1}$ (magenta). Bottom panel: time evolution of $J_z$ at the centre of the current sheet for cases B7, B8 and B9, run at  $\eta = 5 \cdot 10^{-4}$. The simulations are run respectively at $\tau_{_{\operatorname{IR}}} = 5 \cdot 10^{-5}$ (red), $\tau_{_{\operatorname{IR}}} = 5 \cdot 10^{-4}$ (green) and $\tau_{_{\operatorname{IR}}} = 5 \cdot 10^{-3}$ (blue).}
    \label{fig:tau_survey_central_Jz}
\end{figure}

Figure \ref{fig:tau_survey_central_Jz} shows the time variation of current density $J_z$ at $x = 0$, $y = 0$ for the set of simulations at higher $\eta$ (top panel) and at lower $\eta$ (bottom panel). The beginning of reconnection is identiﬁed with the ﬁrst minimum occurring in the current density, when the current sheet is compressed the most by the two initial plasmoids.
The onset of reconnection begins at similar times for all the simulations in the survey, occurring later at the increasing of $\tau_{_{\operatorname{IR}}}$ as ionisation and recombination processes become more important in varying the plasma composition around the $X$-point. For both sets of simulations, the differences between these cases in the initial phase of the coalescence are small as the initial collisional coupling is very weak, and the central current sheet forms in nearly complete absence of collisions.

In the set run at $\eta = 1.5 \cdot 10^{-3}$ (cases B1 to B6 in Table \ref{tab:parameters_2D}), several differences are observed in the reconnection dynamics of each simulation. Case B1, which is run at the lowest $\tau_{_{\operatorname{IR}}} (= 5 \cdot 10^{-6})$, is the only case at higher $\eta$ where secondary plasmoids formation occurs in the central current sheet, as shown by the fluctuation of the gray curve in the top panel of Figure \ref{fig:tau_survey_central_Jz}. At $\alpha_c = 5$ the fluids are weakly coupled, with 0.05 collisions due to happen in a unit of time: this aspect allows the plasma to evolve separately with respect to the neutral fluid, and promotes secondary plasmoids formation. In case B1 the current sheet collapse (identified as the time between $t = 0$ and the first minimum in the current, where the current sheet is compressed the most) occurs over a period $\Delta t \sim 0.9$: at the imposed collisional coupling, we are expecting 1 collision every $\Delta t \sim 17$, therefore we consider the current sheet formation to occur in an almost collisionless regime. The collisional ionisation rate of case B1, initially $5 \cdot 10^{-8}$, do not significantly increase the ion fraction during the first phase of coalescence due to the lack of collisions, therefore plasmoids can form before ionisation can occur. The low ion fraction leads to a more efficient current sheet thinning, as seen in our previous paper\cite{doi:10.1063/5.0032236}. The ejection of the first plasmoid leaves the current sheet unstable, and further smaller plasmoids are produced, leading to turbulent reconnection that efficiently reduces the coalescence time scale. Increasing $\tau_{_{\operatorname{IR}}}$ by one order of magnitude (case B2, red curve in top panel of Figure \ref{fig:tau_survey_central_Jz}), the ionisation rate converts a portion of neutrals large enough to increase the current sheet thickness and prevent the formation of further dynamics. For $\tau_{_{\operatorname{IR}}} \ge 5 \cdot 10^{-5}$ the tearing instability is suppressed, as the ionisation rates is sufficiently high to stabilise the current sheet.

In the simulations run at $\eta = 1.5 \cdot 10^{-3}$ the magnitude of $J_z$ rapidly increases with $\tau_{_{\operatorname{IR}}}$, while the time scale for the plasmoid coalescence becomes considerably shorter. The faster coalescence is proven by the end time of the evolution in the top panel of Figure \ref{fig:tau_survey_central_Jz}, where the merging completion can be identified by the current density reaching positive values (at $t > 5$ for all cases with the exception of case B1, corresponding to the gray curve). The reconnection phase shortens with the thinning of the current sheet. The thinning is faster for cases where $\Gamma_{ion}$ is bigger, as the cooling action of the ionisation potential is proportional to the ionisation rate. Increasing $\tau_{_{\operatorname{IR}}}$, and consequently $\Gamma_{ion}$, reconnection becomes faster and leads to larger current densities, while the current sheet becomes thinner. In perspective, at $\tau_{_{\operatorname{IR}}} > 0.5$ we might expect a regime where the current sheet would be thin enough to promote secondary dynamics again.

In the interval of values $\tau_{_{\operatorname{IR}}} = [5 \cdot 10^{-5}, 5 \cdot 10^{-1}]$, the onset of the tearing instability might be achieved when a smaller diffusivity is chosen for the system, as thinner current sheets would form. For this reason we investigate three PIP cases (B7 to B9 in Table \ref{tab:parameters_2D}) run with $\tau_{_{\operatorname{IR}}} = [5 \cdot 10^{-5}$, $5 \cdot 10^{-3}]$ for a diffusivity $\eta = 5 \cdot 10^{-4}$. Bottom panel of Figure \ref{fig:tau_survey_central_Jz} shows the current density at the centre of the current sheet for the cases at $\tau_{_{\operatorname{IR}}} = 5 \cdot 10^{-5}$ (red), $5 \cdot 10^{-4}$ (green) and $5 \cdot 10^{-3}$ (blue). Secondary plasmoid formation is observed for all three cases, as the smaller resistivity leads to thinner current sheets and promotes the onset of the tearing instability. At the lowest $\tau_{_{\operatorname{IR}}}$ this non-linear dynamics shortens the coalescence time scale, as seen by the comparison of the red curves between top and bottom panel of Figure \ref{fig:tau_survey_central_Jz}.

\begin{figure}[htb]
    \centering
    \includegraphics[width=\columnwidth,clip=true,trim=0.5cm 8.5cm 0.5cm 8cm]{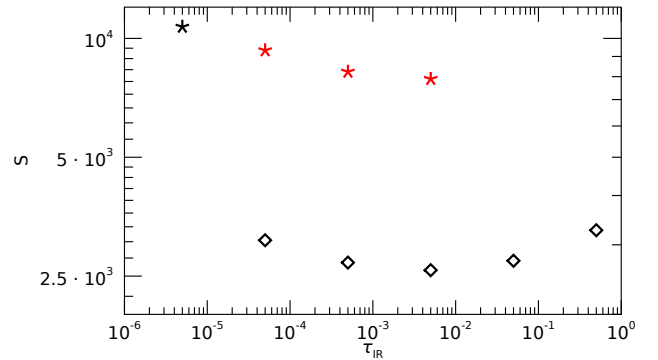}  
    \caption{Lundquist number of the $\tau_{_{\operatorname{IR}}}$ survey PIP simulations (B1 to B9) at the beginning of the reconnection phase. Stars are associated with simulations that develop secondary plasmoids in the central current sheet, diamonds represent simulations that do not have secondary plasmoids. Black symbols refer to the cases run at $\eta = 1.5 \cdot 10^{-3}$, red symbols to the cases run at $\eta = 5 \cdot 10^{-4}$.}
    \label{fig:tau_survey_S_current_minimum}
\end{figure}

Figure \ref{fig:tau_survey_S_current_minimum} shows the Lundquist number of the simulations in the survey, calculated at the beginning of the reconnection phase by using the effective Alfv\'en speed $v_{A,e}$. In the cases at higher $\eta$, $S$ decreases at the increase of $\tau_{_{\operatorname{IR}}}$, increasing again for $\tau_{_{\operatorname{IR}}} > 0.005$.
We identify a threshold for the critical Lundquist number laying in the interval $S = 5.1 \cdot 10^3 - 7.9 \cdot 10^4$. The simulation at the lowest $\tau_{_{\operatorname{IR}}}$ (B1) is the only case at $\eta = 1.5 \cdot 10^{-3}$ where secondary plasmoids are seen to form in the central current sheet. The Lundquist number for this simulation is $1.07 \cdot 10^4$, above the critical Lundquist number. The simulations run at $\eta = 5 \cdot 10^{-4}$ develop plasmoid formation at Lundquist numbers ($S = 9.3 \cdot 10^3$ for B7, $S = 8.2 \cdot 10^3$ for B8 and $S = 7.9 \cdot 10^3$ for B9) that are consistent to the critical numbers found in our previous paper \cite{doi:10.1063/5.0032236}.

\begin{figure*}[htb]
    \centering
    \includegraphics[width=\textwidth,clip=true,trim=0cm 1.5cm 0cm 0cm]{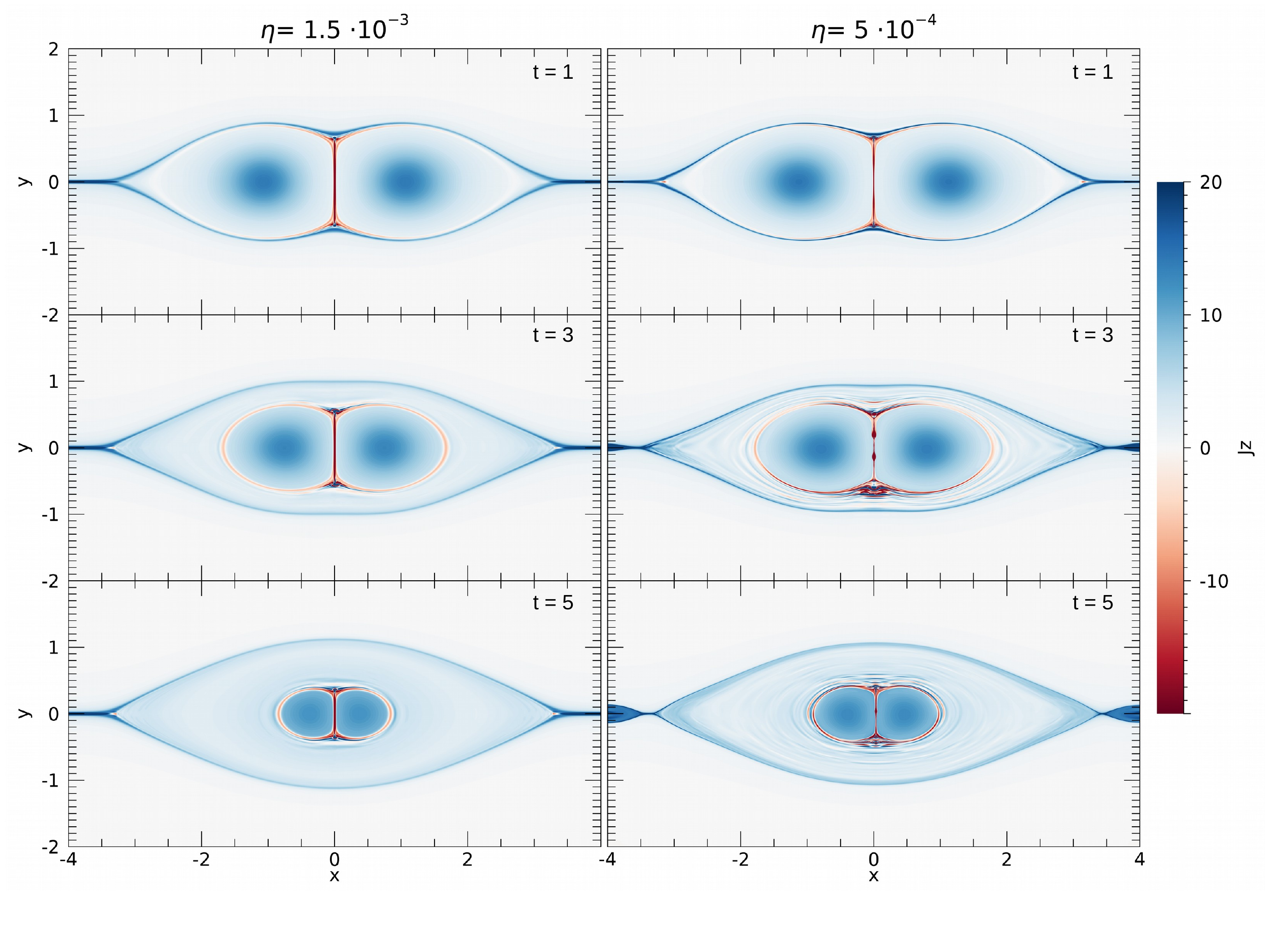} 
    \caption{Comparison of J$_z$ between the PIP cases B4 (left column) and B9 (right column). The frames show the coalescence instability at three different times during the reconnection phase. Times are given in the same non-dimensional unit. In the case at lower $\eta$ (right column) secondary plasmoids form in the central current sheet (right central and bottom panels).}
    \label{fig:plasmoids_eta_difference}
\end{figure*}
At higher $\tau_{_{\operatorname{IR}}}$, despite having a copious production of secondary plasmoids as shown by the the fluctuation in the current density, the coalescence time scale is less affected by plasmoid dynamics. This effect might be explained by the reconnection rate saturation already observed in many 2D MHD simulations\cite{2009PhPl...16k2102B,2012ApJ...760..109L,2017ApJ...848..102T}, whose study revealed that above the critical Lundquist number the reconnection rate becomes almost independent of $S$, following the onset of nonlinear dynamics. Figure \ref{fig:plasmoids_eta_difference} shows a comparison between two PIP cases at $\tau_{_{\operatorname{IR}}} = 5 \cdot 10^{-3}$ that are run at two different $\eta$, for three times ($t = 1, 3$ and $5$). These cases, listed as B4 and B9 in Table \ref{tab:parameters_2D}, are identified by the blue curves in both top and bottom panel of Figure \ref{fig:tau_survey_central_Jz}. As shown by the direct comparison of the coalescing plasmoids size between the two cases, the coalescence proceeds at a similar time scale even though in one case turbulent reconnection occurs. The Lundquist numbers of the two simulations are respectively $S_{B4} = 2.59 \cdot 10^3$ and $S_{B9} = 7.89 \cdot 10^3$: secondary plasmoids in case B9 develop for an $S$ consistent to the Lundquist numbers at which secondary plasmoids form in PIP of our previous paper\cite{doi:10.1063/5.0032236}.
\begin{figure}[htb]
    \centering
    \includegraphics[width=\columnwidth,clip=true,trim=1cm 8.5cm 2cm 8.7cm]{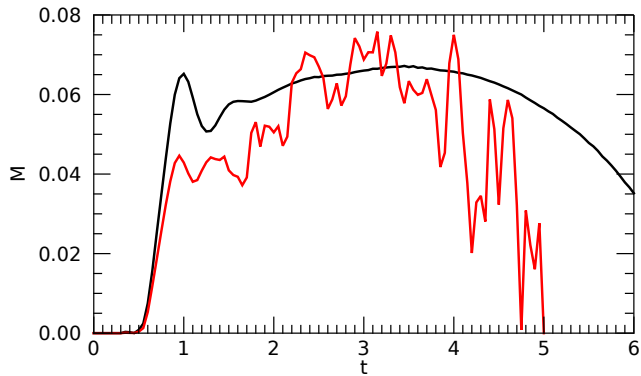}  
    \caption{Time variation of the reconnection rate for the PIP cases B4 (black curve) and B9 (red curve). The large fluctuations in case B9 depend on turbulent reconnection taking place.}
    \label{fig:M_comparison}
\end{figure}
The reconnection rate of both simulations is shown in Figure \ref{fig:M_comparison}. Despite having a difference in the Lundquist number, the two rates evolves in a similar way, with the larger fluctuations occurring in case B9 following the secondary plasmoid dynamics. The mean reconnection rates, calculated between $t = 1$ and $t = 5$, are respectively $M_{B4} = 0.062 \pm 0.004$ and $M_{B9} = 0.05 \pm 0.02$, where the errors are calculated as the standard deviation. The two rates are close, which is consistent with the effect of reconnection rate saturation observed in other works\cite{2009PhPl...16k2102B,2012ApJ...760..109L,2017ApJ...848..102T}. From the Sweet-Parker steady state reconnection model, the change in $\eta$ of a factor of 3 between cases B4 and B9 is expected to result in a small change of $\sim 1/ \sqrt{3}$ in the reconnection rate. Re-scaling the black curve in Figure \ref{fig:M_comparison} by this factor to make a prediction on the reconnection rate at lower $\eta$ we see a large difference of the predicted value with $M_{B9}$: the consistency between $M_{B4}$ and $M_{B9}$ is therefore not due to the small change in $\eta$ leading to similar rates, but on the reconnection rate saturation itself. occurring once the turbulent reconnection is set.
 
\begin{figure}[htb]
    \centering
    \includegraphics[width=\columnwidth,clip=true,trim=2.3cm 7.5cm 3.7cm 8.7cm]{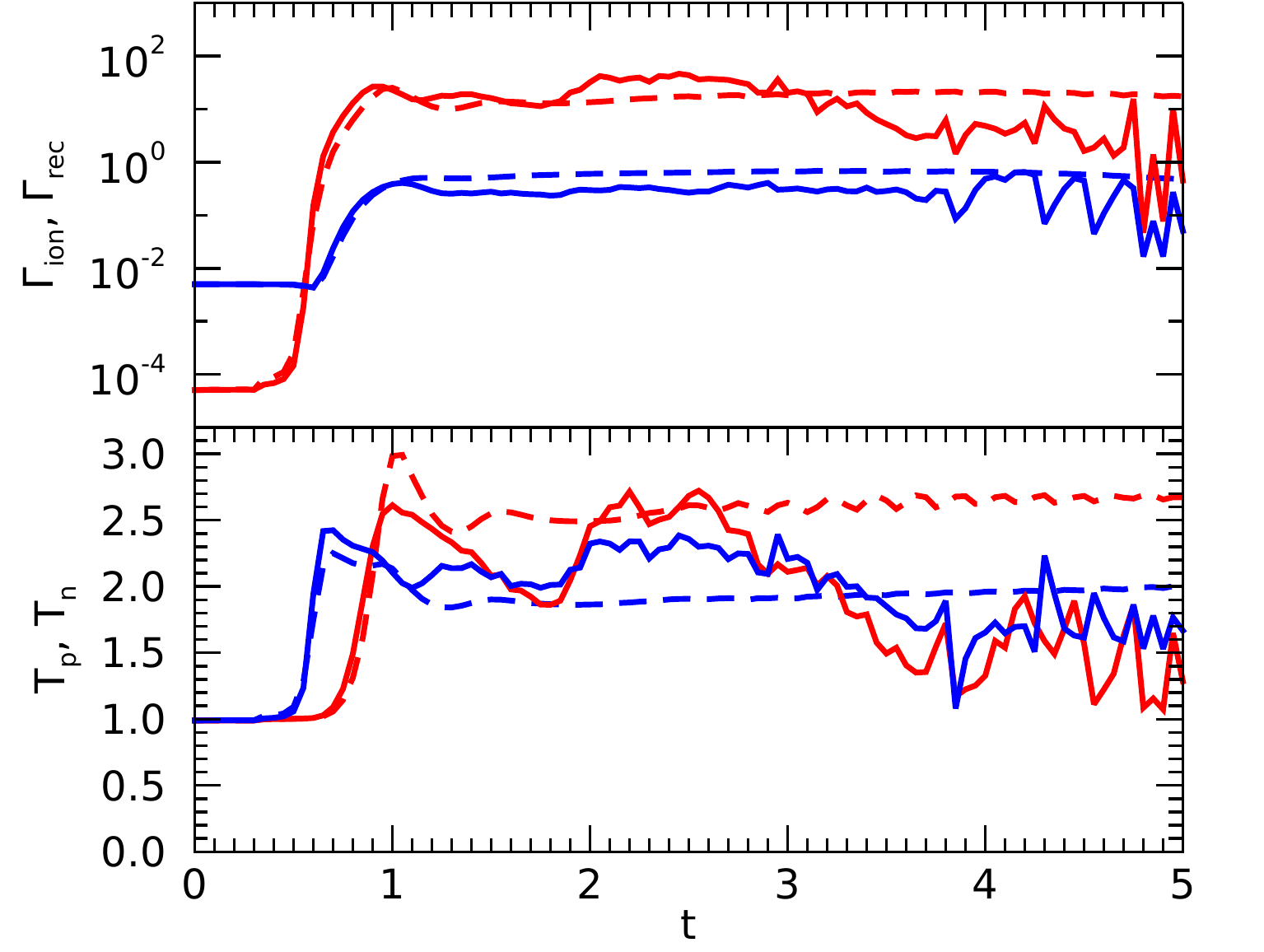} 
    \caption{Top panel: time variation of the mean ionisation and recombination rates inside the current sheet for the PIP case B4 (dashed line, $\eta = 1.5 \cdot 10^{-3}$) and B9 (solid line, $\eta = 5 \cdot 10^{-4}$). Ionisation rates are shown in red, recombination rates are shown in blue. Bottom panel: time evolution of the mean plasma (blue) and neutral (red) temperature inside the current sheet for the same PIP cases B4 (dashed line) and B9 (solid line).}
    \label{fig:temp_rates_eta_comparison}
\end{figure}

The time variation of the mean plasma and neutral temperatures and the mean ionisation and recombination rates inside the current sheet is shown in Figure \ref{fig:temp_rates_eta_comparison} for cases B4 (dashed lines) and B9 (solid lines).
$\Gamma_{ion}$ and $\Gamma_{rec}$ (top panel of Figure \ref{fig:temp_rates_eta_comparison}) increase in both cases in the ideal phase of coalescence following the initial ionisation burst at the formation of the current sheet, and are maintained approximately constant in time in the reconnection phase, with small fluctuations around the average value. The same behaviour is observed in the temperature variation (bottom panel of Figure \ref{fig:temp_rates_eta_comparison}). In case B4  $T_p$ (blue dashed line) is approximately constant over the reconnection phase, while $T_n$ (red dashed line) shows small regular fluctuations at $t > 2.5$. Plasma is colder than the neutral counterpart. This happens because reconnection leads to a high plasma temperature initially, and once ionisation occurs large amounts of plasma energy are lost and the plasma fluid cools. As the ionisation burst happens over a short period of time, the time scale isn't long enough for the plasma temperature to couple to the neutral temperature. In case B4, where laminar reconnection occurs, a balance is maintained with the species entering the current sheet from the inflow and the mean temperatures vary slowly, thus leading to relatively steady ionisation/recombination rates.

The fluctuations increase with the onset of the tearing instability in the current sheet as shown by the solid lines in the top panel of Figure \ref{fig:temp_rates_eta_comparison}. In case B9, after an increase of temperature at the current sheet initial formation, larger fluctuations appear at the onset of tearing instability. At $t>1$ the plasma temperature (solid blue line) is larger than case B4, while $T_n$ is smaller. The larger $T_p$ is the result of the larger compression of the current sheet which is thinner in case B9. After $t \sim 3.5$ the two mean temperatures in the current sheet reach an equilibrium and fluctuate around a similar average value. The mean temperatures at $t > 3.5$ are respectively $T_p = 1.7 \pm 0.2$ and $T_n = 1.5 \pm 0.2$, where the ranges are calculated as the standard deviation of the measures. At later stages there is also a general decrease of both $\Gamma_{ion}$ and $\Gamma_{rec}$ in case B9, while in case B4 the rates remain constant. The large fluctuations in presence of secondary plasmoids dynamics depend on the different mechanism of expulsion of the plasma from the unstable current sheet. The neutral temperature decreases at later times in correspondence of secondary plasmoids formation - and consequently does $\Gamma_{ion}$ - and equals the plasma temperature, whose magnitude is comparable to the case without tearing instability (B4).

In this Section we investigate the relative importance of $\Gamma_{ion}$, $\Gamma_{rec}$ to the collisional coupling $\alpha_c$ on determining the type of reconnection during plasmoid coalescence. Because of the nature of the coalescence instability and the large compression of the current sheet in a small time interval, very small ionisation/recombination rates ($\tau_{_{\operatorname{IR}}} \sim 5 \cdot 10^{-5}$) are capable to affect the reconnection dynamics, as they are very sensible to temperature changes. Turbulent reconnection in weakly coupled plasmas is promoted either for nearly negligible ionisation/recombination rates ($\tau_{_{\operatorname{IR}}} \sim 5 \cdot 10^{-6}$) or by reducing $\eta$, which directly affects the Lundquist number.

\subsection{Survey on the ion fraction}
\label{sec:ion_fraction}

We investigate the changes in the coalescence dynamics of the IRIP cases following a variation of the initial ion fraction $\xi_p$. The parameter $\xi_p$ depends directly on the reference temperature $T_0$, selected at the beginning of calculation. Therefore, we evaluate changes in the coalescence instability by progressively increase $T_0$ by 1000 K. We compare four calculations, listed as C1, B2, C2 and C3 in Table \ref{tab:parameters_2D}, where we vary the reference temperature in the range $T_0 = [9855, 12855]$ K.

\begin{figure}[htb]
    \centering
    \includegraphics[width=\columnwidth,clip=true,trim=0cm 0cm 0cm 0cm]{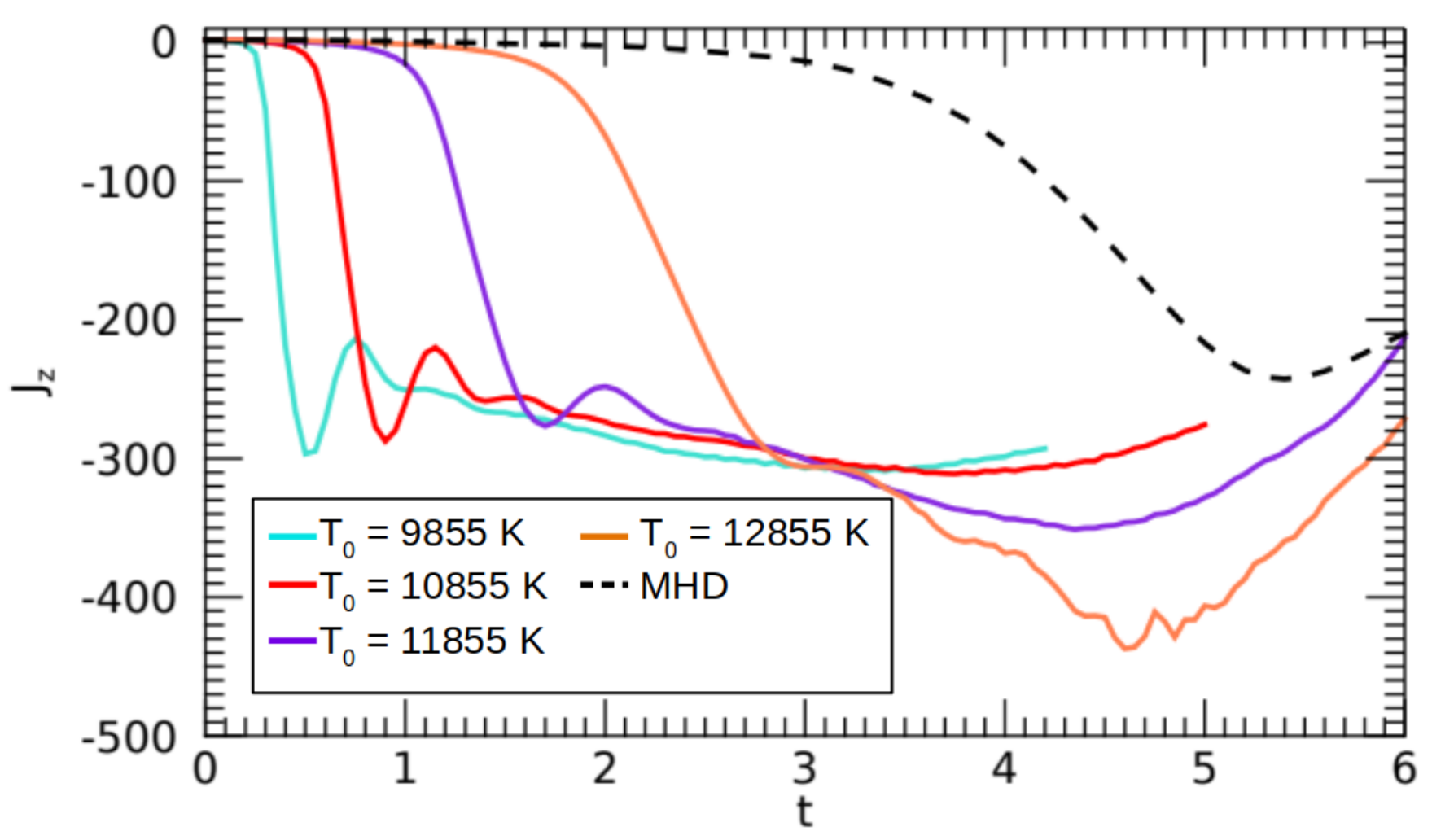}
    \caption{Time evolution of current density $J_z$ at the centre of the current sheet for four PIP cases (C1,C2,C3 and B2). The initial reference temperature $T_0$ is respectively $T_0 = 9855$ K, corresponding to an initial $\xi_p = 2 \cdot 10^{-3}$ (turquoise), 10855 K, corresponding to $\xi_p = 10^{-2}$ (red), 11855 K, corresponding to $\xi_p = 4 \cdot 10^{-2}$ (purple) and 12855 K, corresponding to $\xi_p = 10^{-1}$ (orange). The current density of an MHD case (M1) is included as comparison to the limit $T_0 \rightarrow \infty$.}
    \label{fig:Jz_temp}
\end{figure}

Figure \ref{fig:Jz_temp} shows the time variation of $J_z$ at the centre of the current sheet for the four IRIP cases. The changes in $T_0$ correspond to an ion fraction variation in the range $\xi_p = [2 \cdot 10^{-3}, 10^{-1}]$. Note that in all cases we consider a medium that is initially dominated by neutrals.

The coalescence time scale is drastically reduced at the decrease of temperature, and consequently ion fraction. The time scale shortening, involving both ideal phase (initial attraction of the plasmoids to each other) and reconnection phase, is explained by the variation of the effective Alfv\'en speed, which scales as $1/ \sqrt{\xi_p}$. The increased Alfv\'en speed allows more flux to enter the reconnection region, hence feeding the reconnection process and accelerating it.

In terms of the observed trend in the development of the current density magnitude, this survey reproduces the results of our previous paper's study on $\xi_p$\citep{doi:10.1063/5.0032236}. Going towards higher $\xi_p$, where the magnetic forces are felt by larger portions of the ﬂuid, the coalescence time scale tends to the MHD case. The reconnection rate increases slightly as $\xi_p (0)$ increases. This trend, shown by the values of mean reconnection rate collected in Table \ref{tab:temperature_cases}, is opposite of what was observed in our previous paper\citep{doi:10.1063/5.0032236}. 
\begin{table}
\caption{Initial ion fraction $\xi_p (0)$, mean reconnection rate $M$ and Lundquist number $S$ of cases C1, B2, C2 and C3, run for the survey on the ion fraction. Errors on the reconnection rate are calculated as the standard deviation. \label{tab:temperature_cases}}
\begin{ruledtabular}
\begin{tabular}{cccc}
 ID & $\xi_p (0)$ & M & S \\
 \hline
 C1 & $2 \cdot 10^{-3}$ & 0.063 $\pm$ 0.005 & $2.7 \cdot 10^3$ \\
 B2 & $10^{-2}$ & 0.063 $\pm$ 0.005 & $3.1 \cdot 10^3$ \\
 C2 & $4 \cdot 10^{-2}$ & 0.067 $\pm$ 0.008 & $5.1 \cdot 10^3$ \\
 C3 & $10^{-1}$ & 0.08 $\pm$ 0.01 & $4.9 \cdot 10^3$
\end{tabular}
\end{ruledtabular}
\end{table}
\begin{figure}[htb]
    \centering
    \includegraphics[width=\columnwidth,clip=true,trim=0.6cm 4cm 0.6cm 5cm]{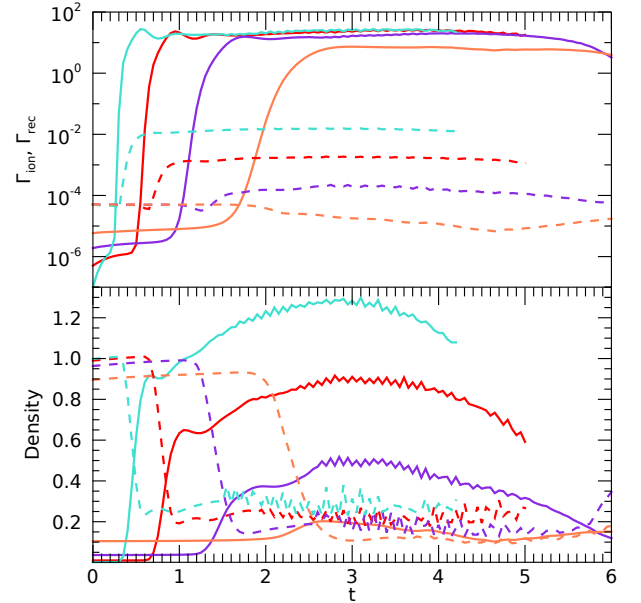}  
    \caption{Top panel: time variation of the mean ionisation (solid lines) and recombination rates (dashed lines) averaged over the current sheet for cases C1 (turquoise), B2 (red), C2 (purple) and C3 (orange). Bottom panel: time variation of the mean values of $\rho_p$ (solid lines) and $\rho_n$ (dashed lines) inside the current sheet for for cases C1 (turquoise), B2 (red), C2 (purple) and C3 (orange). The oscillations observed in the density are related to the oscillatory motion of the current sheet.}
    \label{fig:temperature_survey_parameters}
\end{figure}
The reversal in trend might be explained by examining both the variation in plasma and neutral densities across the current sheet, and the average ionisation and recombination rates. Figure \ref{fig:temperature_survey_parameters} shows the mean values of $\Gamma_{ion}$ and $\Gamma_{rec}$ (top panel) and of $\rho_p$ and $\rho_n$ (bottom panel) across the current sheet for the four cases presented in the survey. The oscillations of the average value of $\rho_p$ and $\rho_n$ can be associated to the oscillatory motion of the current sheet.

The compression of the current sheet by the merging plasmoids results in heating that leads to a burst in ionisation, observed in particular at the formation of the current sheet. At the increase of the initial $\xi_p$, both $\Gamma_{ion}$ and $\Gamma_{rec}$ decrease inside the current sheet, as shown in the top panel of Figure \ref{fig:temperature_survey_parameters}. This is due to the larger $\rho_p$ at the centre of the structure, as the plasma pressure prevents the current sheet to thin more, hence showing a smaller temperature change. When the ionisation rate is small it leads to a limited ionisation, which also turns into a lesser thinning of the current sheet as the cooling action of the ionisation potential is reduced. The lower cooling affects the recombination rate, whose decrease leads to fewer neutrals forming and leaving the current sheet.

Having an initial lower $\xi_p$, and consequently a larger ionisation rate due to the initial current sheet thinning, both $\rho_p$ and $\rho_n$ increase more inside the current sheet, the latter due to the larger cooling factor provided by the ionisation potential. Although the coalescence is faster for lower $\xi_p (t = 0)$, the local increase in density leads to a comparable reconnection rate for all the cases in this survey.

The change of the effective Alfv\'en speed and plasma density affects the Lundquist number, which in turn extends the time scale of the reconnection phase. The values for the Lundquist number calculated at the beginning of the reconnection phase are shown in Table \ref{tab:temperature_cases}, and it is shown that the Lundquist number also slightly increases in the interval. The value of $S$ for case C3 ($\xi_p = 10^{-1}$) is consistent with our previous study without ionisation and recombination\citep{doi:10.1063/5.0032236}. At lower $\xi_p (t = 0)$ the Lundquist number is generally lower than our previous work, but this is due to the increase in plasma density that comes from the ionisation at the centre of the current sheet.

In general, the evolution of the coalescence instability observed from changes in the current density follows the same trend as the NIR cases previously studied \citep{doi:10.1063/5.0032236}. However, ionisation/recombination rates act on both the reconnection rate $M$ and the Lundquist number $S$. Both parameters vary less than the NIR simulations in the same $\xi_p(0)$ interval.

\section{Discussion}
\label{sec:discussion}

Plasmoid coalescence is a very important process for promoting fast magnetic reconnection in many reconnecting systems: in fact, plasmoids allow the shortening of the time scale for explosive events by acting on the size of reconnecting current sheets. It is not entirely clear about how coalescence develops in a partially ionised plasma and in particular how it is affected by the charge-neutral species interaction. In this work we have extended our previous study on the coalescence instability in partially ionised plasmas \citep{doi:10.1063/5.0032236} by including ionisation, recombination and optically thin radiative losses in the two-fluid coupling terms. The goal of our update is to model a more realistic coupling between different particle species during coalescence. We have observed several changes in the development of this instability that are related to the newly included two-fluid physics:
\begin{enumerate}
 \item Ionisation and recombination have a stabilising effect on the current sheet as they are responsible for thickening it, but still lead to a faster reconnection in PIP when compared to MHD cases at the same bulk density.
 \item The formation of the central current sheet begins at similar times for all PIP cases at the same initial collisional coupling, with or without ionisation and recombination, ionisation potential and background heating. This is because in the first phase of coalescence the rates are in equilibrium, and the temperature on which they depend does not show sharp variation. During the reconnection phases, where the temperature increases, ionisation/recombination effects dominate the reconnection dynamics, and are responsible for the type of reconnection that develops in the current sheet.
 \item The internal structure of secondary plasmoids is slightly altered under the action of ionisation and recombination: a larger part of their structure is in magnetohydrostatic equilibrium, while the external region characterised by a force-free equilibrium is reduced to a thin annulus. However, the secondary plasmoid dynamics inside the current sheet is unchanged when compared to NIR cases \citep{doi:10.1063/5.0032236}.
 \item Small initial ionisation/recombination rates ($\sim$ five orders of magnitude smaller than the collision rates) in weakly ionised plasmas can still lead to an efficient stabilisation of the current sheet against the tearing instability in the reconnection phase. This happens because the temperature increases locally enhance the rates by several orders of magnitude inside the current sheet, thus leading to a substantial increase in $\rho_p$ inside the current sheet. 
 \item When varying the initial ion fraction, the general development of the coalescence is unchanged from NIR cases, but ionisation and recombination influence reconnection rate and Lundquist number because they increase locally the plasma density, thus affecting the value of the effective Alfv\'en speed.
\end{enumerate}
In addition to the results that are directly dependent on the presence of ionisation/recombination rates, we also find that in cases of laminar reconnection the current sheet develops an oscillatory behaviour in the outflow that is independent of partial ionisation effects, and correlated to the current sheet thickness (see Figure \ref{fig:vel_div}).

From our findings, the stabilisation factor in (1) provides interesting consequences for a recent paper\citep{2015PASJ...67...96S} that had investigated fractal reconnection in partially ionised plasmas in strongly, intermediate and weakly coupled regimes. As changes in the ionisation level directly affects the current sheet evolution, then the time scales for fractal tearing previously identified\citep{2015PASJ...67...96S} will be largely affected by the presence of ionisation-recombination processes. The local dynamics observed in (1) and (2) is also consistent with the ionisation bursts found in previous studies on magnetic reconnection in multi-ﬂuid partially ionised plasmas \citep{2018ApJ...868..144N,2012ApJ...760..109L,2013PhPl...20f1202L,2015ApJ...805..134M,2018ApJ...852...95N,2018PhPl...25d2903N}.

We conclude with the implications of our findings for chromospheric reconnection. The chromospheric rates of ionisation and recombination are typically\cite{Carlsson_2002, 2021A&A...645A..81S} in the range $10^{-5} - 10^{-3}$ s$^{-1}$, between 2 and 7 orders of magnitude smaller than the typical chromospheric neutral-ion collision frequency ($10^{-1} - 10^{2}$ s$^{-1}$) \cite{2005A&A...442..1091L,2012ASPC..463..281K,2019A&A...627A..25P}. Our study, where rates are set consistent with the chromospheric values, demonstrates that ionisation and recombination rates are large enough to suppress the onset of fractal coalescence and small-scale dynamics, as they act on current sheets properties. However, multi-fluid physics is still capable to promote fast reconnection, hence explaining the short time scales of chromospheric explosive events. A process that is not accounted for in this work is the interaction of the plasma with the photospheric radiation field. While collisional ionisation by electrons is important for heavier elements, photo-ionisation is the predominant process for ionising hydrogen in the chromosphere\citep{1985irss.rept..213G} and leads to a hydrogen ionisation rate\citep{1989A&A...225..222V,1998A&A...333.1069P} $\Gamma_{ion}^{ph} \sim 1.4 \cdot 10^{-2}$ s$^{-1}$ that can become orders of magnitude bigger than collisional ionisation by electrons\citep{1967ApJS...14..207L,1998A&A...333.1069P} $\Gamma_{ion} \sim 7.8 \cdot 10^{-5}$ s$^{-1}$. The inclusion of photo-ionisation would impose changes in the balance between ions and neutrals as function of the atmospheric temperature, therefore it will be subject for further updates of our model.

\acknowledgments

The authors are grateful to Dr. N. Nakamura for the inspiration of his original work on this problem that lead to this study. AH and BS are supported by STFC Research Grants No. ST/R000891/1 and ST/V000659/1.

This work used the DiRAC@Durham facility managed by the Institute for Computational Cosmology on behalf of the STFC DiRAC HPC Facility (www.dirac.ac.uk). The equipment was funded by BEIS capital funding via STFC capital grants ST/P002293/1, ST/R002371/1 and ST/S002502/1, Durham University and STFC operations grant ST/R000832/1. DiRAC is part of the National e-Infrastructure.

\section*{Data availability}

The data that support the findings of this study are available from the corresponding author upon reasonable request. The (P\underline{I}P) code is available at the following url: \url{https://github.com/AstroSnow/PIP}. Details of the code and equations are available in A. Hillier, S. Takasao \& N. Nakamura, A\&A 591, A112 (2016)\citep{2016A&A...591A.112H}. A copy of the code version used in this work and the initial condition of the reference simulations can be found in a dedicated repository provided by G. Murtas: \url{https://github.com/GiuliaMurtas/PIP.git}.

\bibliography{biblio}

\end{document}